\documentclass[aps,twocolumn,superscriptaddress,showpacs,amssymb,prl,floatfix,longbibliography]{revtex4-1}

\usepackage{graphicx,color}
\usepackage{dcolumn}
\usepackage{amsmath}
\usepackage{amssymb}
\usepackage[]{hyperref}
\hypersetup{colorlinks=true,linkcolor=blue,citecolor=blue,urlcolor=blue,pdfpagemode=UseNone}

\newcommand{\slrrtext}  {spin-lattice-relaxation rate}
\newcommand{\slrr}      {$T_1^{-1}$}

\newcommand{\celacoinx}     {Ce$_{1-x}$La$_x$CoIn$_5$}
\newcommand{\celacoin}     {Ce$_{0.243}$La$_{0.757}$CoIn$_5$}

\begin{document}

\thispagestyle{myheadings}

\title{Site Specific {K}night Shift Measurements of the Dilute {K}ondo lattice System {Ce$_{1-x}$La$_x$CoIn$_5$}}

\author{M. Lawson}

\author{B. T. Bush}

\author{A. C. Shockley}

\affiliation{Department of Physics, University of California, Davis, California 95616, USA}

\author{C. Capan}
\affiliation{Department of Physics and Astronomy, Washington State University, Pullman WA 99164}

\author{Z. Fisk}
\affiliation{Department of Physics, University of California, Irvine, California 92697, USA}

\author{N. J. Curro}

\affiliation{Department of Physics, University of California, Davis, California 95616, USA}

\date{\today}

\begin{abstract}
$^{115}$In Nuclear magnetic resonance data are presented for a series of Ce$_{1-x}$La$_x$CoIn$_5$ crystals with different La dilutions, $x$.  Multiple In(1) sites associated with different numbers of nearest-neighbor cerium atoms exhibit different Knight shifts and spin lattice relaxation rates.  Analysis of the temperature dependence of these sites reveals both an evolution of the heavy electron coherence as a function of dilution, as well as spatial inhomogeneity associated with a complete suppression of antiferromagnetic fluctuations in the vicinity of the La sites.  Quantum critical fluctuations persist within disconnected Ce clusters with dilution levels up to 75\%, despite the fact that specific heat shows Fermi liquid behavior in dilute samples.
\end{abstract}



\maketitle

\section{Introduction}

Heavy fermion compounds exhibit a broad spectrum of novel correlated electron behavior, including unconventional superconductivity and quantum critical phenomena \cite{StewartHFreview,ColemanQCreview}. These materials consist of a lattice of localized $f$-electrons that interact with a sea of itinerant conduction electrons. {For a single impurity the conduction electrons screen the $f$-site below the Kondo temperature, $T_K$, forming a spatially extended singlet state \cite{hewson}.  The situation is considerably more complex for two or more $f$-sites: depending on the relative size of the Kondo and RKKY couplings the $f$-moments can either order antiferromagnetically, or are quenched via Kondo screening \cite{JonesVarmaPRL1987,*VarmaJonesTwoKondoPRL,HirschTwoKondoPRB,doniach,OgataDilutionPRB2010,ZhuCoupledImpurities2011}.
In a fully occupied lattice these competing ground states give rise to a quantum phase transition between these two extremes, where strong fluctuations are responsible for a breakdown of conventional Fermi liquid theory} \cite{YRSnature,ColemanHFdeath}. The microscopic physics in this regime is poorly understood, and key open questions are whether the conduction electrons screen each $f$-moment individually or collectively across multiple sites, and whether the screening is enhanced or suppressed by the second $f$-site \cite{OgataDilutionPRB2010}. There have been few experimental studies of the two-impurity Kondo problem \cite{STM2KondoNature2011}, and as a result there remain several question about the relevant temperature and length scales of the ground state. Here we report nuclear magnetic resonance (NMR) data in \celacoinx, in which the La serves to dilute the lattice of $f$-moments in a prototypical Kondo lattice system close to a quantum phase transition \cite{CeCoIn5discovery,romanQCPCeCoIn5}.  Our results indicate that the heavy electron coherence becomes spatially inhomogeneous and is  suppressed locally in regions with no $f$ sites in a diluted lattice, yet quantum critical fluctuations persist in disconnected clusters.

\begin{figure}
\includegraphics[width=\linewidth]{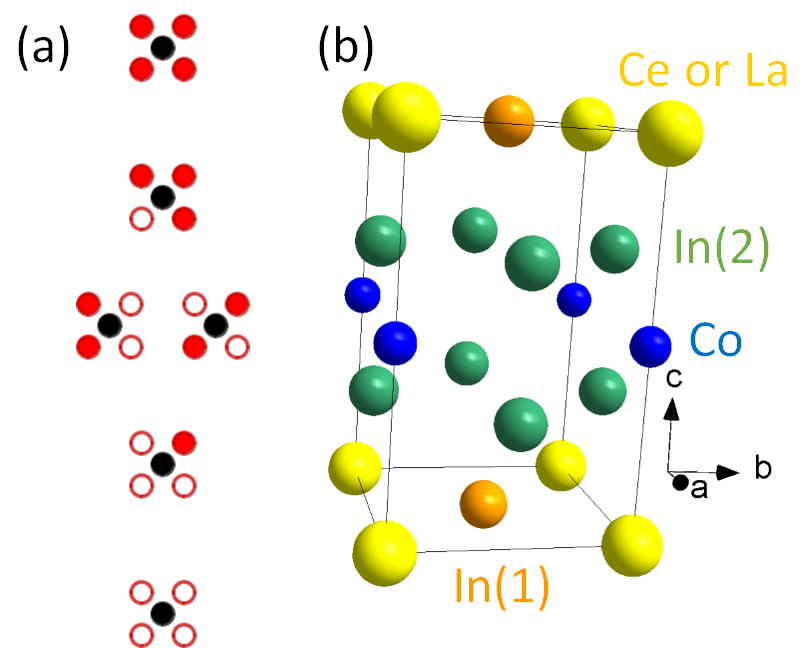}
\caption{\label{fig:unitcell} (color online) (Left) The various combinations of La (empty) and Ce (red) sites surrounding the In(1) (black) site in doped Ce$_{1-x}$La$_x$CoIn$_5$. (Right) The unit cell for pure CeCoIn$_5$.}
\end{figure}

{La dilution is a powerful technique to probe intersite interactions between f-sites in Kondo lattices. }
Replacing the $4f^1$ electron of the Ce$^{3+}$ with a $4f^0$ configuration of La$^{3+}$ removes the local moment without changing the conduction electron count. For sufficiently large La doping, the remaining isolated $4f^1$ Ce moments behave independently and their screening is described by single-ion Kondo physics.
{For random substitutions at La doping levels beyond the percolation limit the lattice will break up into disconnected clusters of $f$-sites with a well defined  size distribution \cite{Percolation3D,*Percolation2Dpart2}. Several years ago pioneering work in Ce$_{1-x}$La$_x$Pb$_3$ \cite{CePb3DilutionPRL} and  \celacoinx \cite{NakatsujiFisk} revealed very different behaviors as a function of $x$. In the former, $T_K$ was observed to be independent of La concentration, suggesting that intersite couplings between the f-sites is negligible. In the latter a new high temperature coherence temperature scale, $T^*\sim 20 T_K$, emerges. Whether $T^*$ represents a renormalized $T_K$ due to differences in the Kondo exchange intergral (a local  effect), or a new energy scale driven by \emph{intersite} couplings remains unclear. Nevertheless,  $T^*$ is clearly evident in various experimental probes  \cite{YangPinesNature,Aynajian2012}, and this observation has laid the foundation of the phenomenological two-fluid description of partially localized f-moments coexisting with an itinerant heavy-electron fluid \cite{NPF,YangPinesPNAS2012}. In order to shed light on the origin of high temperature lattice coherence scale, it is instructive to investigate the spatial dependence of the spin fluctuations and energy scales in a dilute system.  Here we report studies of several different La concentrations in which the $f$-electron clusters continue to exhibit an unusually large coherence temperature and spin fluctuations characteristic of the undoped system. This surprising result reveals that the heavy fermion state is inhomogeneous and suggests that intersite interactions are restricted to nearest neighbors.}

\begin{figure}
    \centering
    \includegraphics[width=\linewidth]{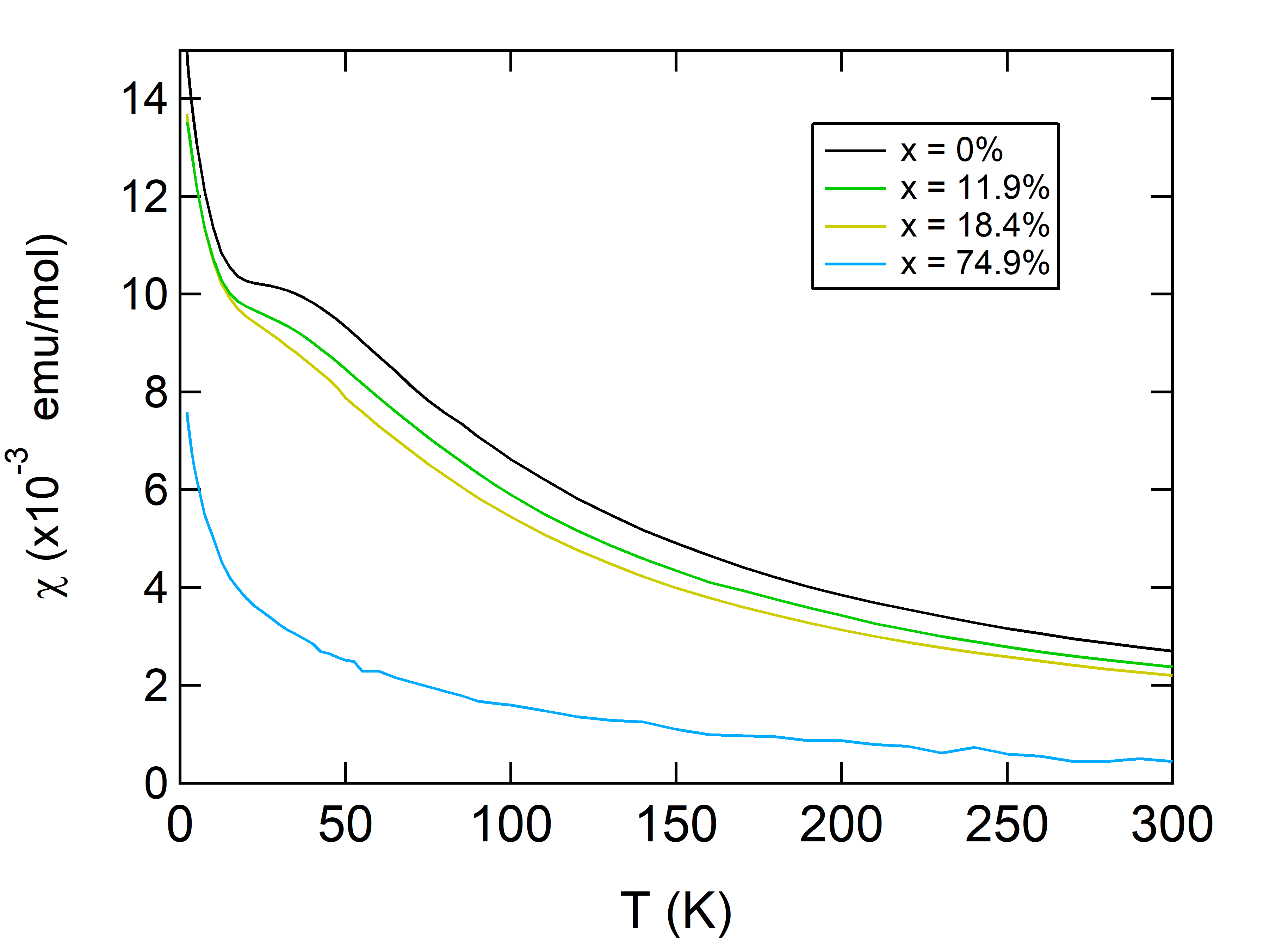}
    \caption{Bulk magnetic susceptibility of the crystals (per formula unit) for field oriented along the $c$-axis.}
    \label{fig:chi}
\end{figure}

NMR is an ideal probe of the local spin correlations that emerge in a Kondo lattice because it provides direct information about $T^*$ and the heavy electron fluid \cite{Curro2009,ShirerPNAS2012}. In CeCoIn$_5$, the In(1), In(2) and Co Knight shifts, $K$, are  proportional to the bulk magnetic susceptibility, $\chi$, for $T>T^*$, but  exhibit a strong Knight shift anomaly below this temperature \cite{Curro2001}. This anomaly originates from the different hyperfine couplings between the nuclear spins and both the conduction electron spins, $S_c$, and the local moment spins, $S_f$. { As a result one can extract detailed information about the three correlation functions $\chi_{\alpha\beta} \sim \langle S_{\alpha} S_{\beta}\rangle$ ($\alpha,\beta = c,f$) by measuring both $K$ and $\chi$ independently \cite{ShirerPNAS2012}.
The spin-lattice relaxation rate, $T_1^{-1}$, probes the spin fluctuations of the local moments and the heavy electron fluid \cite{Yang2009a}.  Here we report $K$ and $T_{1}^{-1}$ for the In(1) site for single crystals with $x =12\%$, 18\% and 75\%.
Both $T^*$ and the magnitude of the heavy electron susceptibility are suppressed with dilution, however the temperature dependence of $T_1^{-1}$ is unaffected by dilution, suggesting that the spin fluctuations persist in disconnected clusters of Ce sites.

\section{Sample Characterization}

Crystals of \celacoinx\ with different nominal La concentrations were synthesized  via flux methods as described in Ref. \onlinecite{NakatsujiFisk}.  The magnetic susceptibilities were measured using a SQUID magnetometer, as shown in Fig. \ref{fig:chi}. Because the La is non-magnetic, the magnetization is dominated by the Ce and the high temperature susceptibility scales as $\chi_{x}(T) \approx (1-x)\chi_0(T)$ \cite{NakatsujiFisk}.  The concentration $x$ was determined by plotting $\chi_{x}$ versus $\chi_0$ and performing a linear fit to the high temperature regime ($T\gtrsim 100$ K).   Based on this analysis we find $x = 11.9 \pm 0.2$\%, 18.4$\pm 0.1$\%, and $74.9 \pm 0.3$\%.  An independent wavelength dispersive spectroscopy (WDS) microprobe analysis on the third sample indicated $x=75.7\pm 1.2$\%.

\section{NMR Spectra}

\begin{figure}
\centering
    \includegraphics[width=\linewidth]{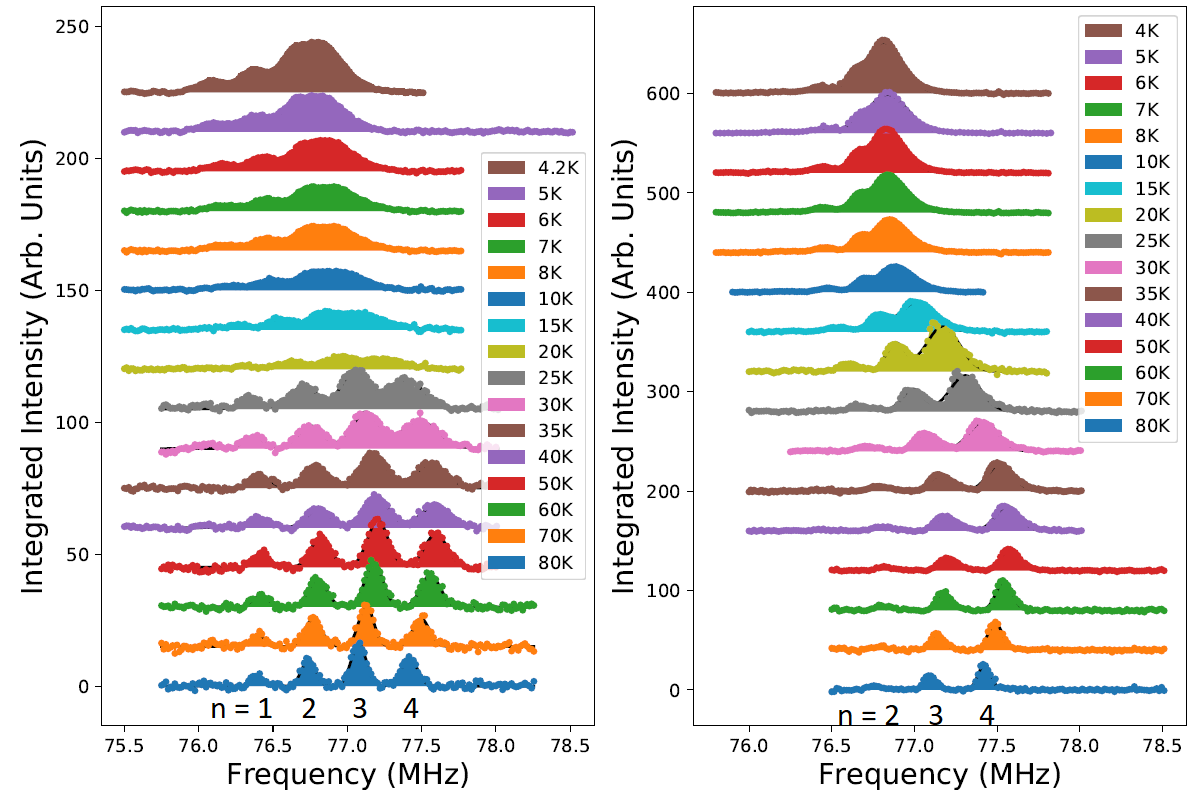}
    \caption{$^{115}$In(1) NMR spectra of the $s=-1$ transition in \celacoinx\ for $x = 18.4\%$ (left) and $x=11.9\%$ (right).  Note the non-monotonic behavior below $T^*\sim 60$K, where the spectra shift to lower frequency.}
    \label{fig:x124_spect}
\end{figure}

Each crystal was oriented with the $c-$axis parallel to a magnetic field $H_0=11.7286$ T and NMR spectra were acquired as a function of frequency and temperature by integrating the spin echo signals.   In this orientation, there are four crystallographically distinct NMR active sites:  $^{139}$La ($I=7/2$), $^{59}$Co ($I=7/2$), and two $^{115}$In ($I=9/2$) sites \cite{Curro2001}. Here, we focus only on the La and the In(1) because the Co and the In(2) spectra are broadened by the disorder.  \slrr\ was determined by fitting the magnetization inversion recovery. Representative spectra of the In(1) site are shown in Fig. \ref{fig:x124_spect}.

The In(1) nuclear spin Hamiltonian is given by: $\mathcal{H} = \gamma\hbar\hat{I}_zH_0 + \frac{h\nu_{cc}}{6}[3\hat{I}_z^2-\hat{I}^2] + \mathcal{H}_{\rm hf}$,  where $\gamma=0.93295$ kHz/G is the gyromagnetic ratio,  $\hat{I}_{\alpha}$ are the nuclear spin operators, $\nu_{cc}$ is the component of the electric field gradient (EFG) tensor along the $c$-direction,  and $\mathcal{H}_{\rm hf}$ is the hyperfine interaction between the In nuclear spins and the electron spins, which gives rise to the Knight shift, $K$ \cite{CPSbook}.
For the In(1) ($I=9/2$) in this configuration, the resonance frequencies are given by: $f = \gamma H_0(1+K) + s \nu_{cc}$, where $s = -4,-3, \cdots, +4$ corresponding to a central transition ($s=0$) and eight satellites, and $K$ is the Knight shift.  The spectra shown in Fig. \ref{fig:x124_spect} correspond to the $s=-1$ satellite.

\begin{figure*}
\centering
    \includegraphics[width=\linewidth]{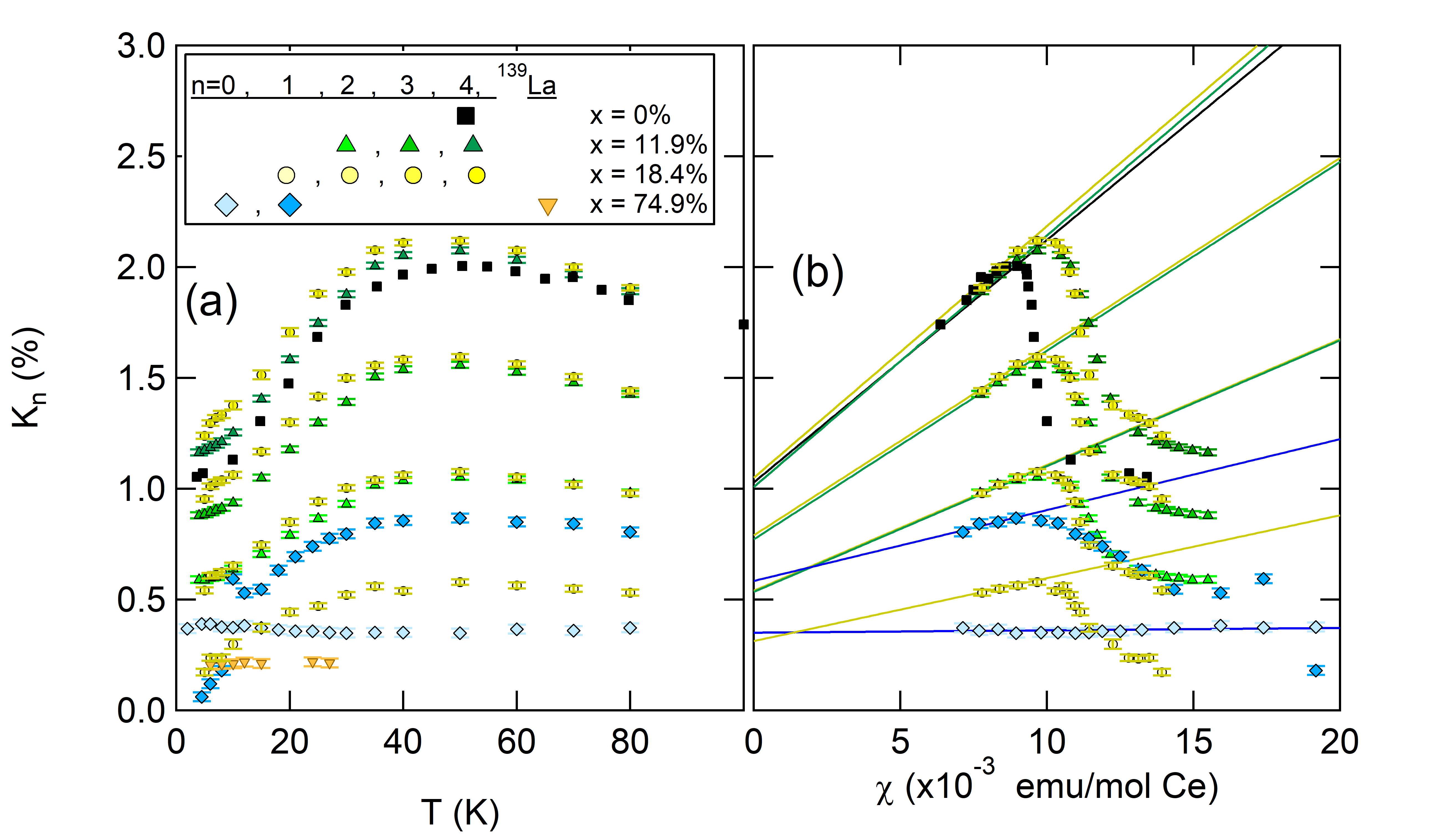}
    \caption{(a) The Knight shift of the In(1) sites, $K_n$ for all measured sites, $n$ and La concentrations, $x$, as a function of temperature. The orange triangles ($\blacktriangledown$) are the Knight shift of the $^{139}$La site for the $x=75\%$ sample. (b) Knight shift versus bulk susceptibility, $\chi/(1-x)$, normalized by the number of Ce atoms per unit cell.  Solid lines are fits to the high temperature points as described in the text. Symbols are identical to those in panel (a). Data for $x=0$ is reproduced from Ref. \cite{Curro2001}.}
    \label{fig:KvsT}
\end{figure*}

Multiple peaks are evident in the spectra, which correspond to sites with different local Knight shift and EFG parameters.  Doping creates variations in the local environment of the In nuclei, which have different numbers of nearest neighbor Ce sites.  As shown in Fig. \ref{fig:unitcell}, the In(1) site has $n = 0, 1, 2, 3$ or $4$ Ce neighbors, and therefore there are potentially six distinct sites in a sample with a finite La concentration, although the two $n=2$ sites may be indistinguishable.  The relative populations, $P_n$ of an In(1) site with $n$ nearest Ce neighbors randomly distributed are given by the binomial distribution: $P_0=x^4$, $P_1=4 x^3 (1 - x)$, $P_2 =6 x^2 (1-x)^2$, $P_3=4 x (1 -x)^3$, and $P_4=(1-x)^4$.  The spectra of different satellites, $s$, reveal the same series of peaks, with identical frequency spacing between the peaks.  We thus conclude that the peaks correspond to different Knight shifts, and the local EFG variations are minor compared with the local hyperfine field variations \cite{Shockley2012}. Each spectrum was fit to a sum of multiple Gaussians to extract the Knight shifts, shown as a function of temperature for various dopings in Fig. \ref{fig:KvsT}(a).

\section{Knight Shift Analysis}

With the exception of the lower peak in the spectra of the $x=75\%$ sample, the shifts of the different sets of peaks shown in Fig. \ref{fig:KvsT}(a) appear to scale with one another with a common temperature dependence.  We postulate that the different peaks observed in Fig. \ref{fig:x124_spect} arise from different numbers, $n$,  of nearest-neighbor Ce atoms, with different Knight shifts, $K_n$, as demonstrated in Fig. \ref{fig:unitcell}. The temperature-independent behavior of the lower peak for the $x=75\%$ sample (Fig. \ref{fig:KvsT}) corresponds to a site with $n=0$, i.e., zero nearest-neighbor Ce sites.  Fig. \ref{fig:KvsK} displays $K_{n}$ versus $K_4$, which reveals linear behavior for all data sets with slopes equal to $n/4$ for temperatures $T\gtrsim 25$ K. This behavior indicates that $K_n\sim n$, and that the dominant contribution to the shift arises from the transferred coupling to the Ce spins, $\mathbf{S}_f$. As seen in Fig. \ref{fig:KvsT}, the Knight shift decreases with the temperature below about 40K for all sites in all samples as the local moments get screened and the heavy quasi-particles form. This is Knight shift anomaly originally reported in \cite{Curro2004}.

\begin{figure}
\centering
    \includegraphics[width=\linewidth]{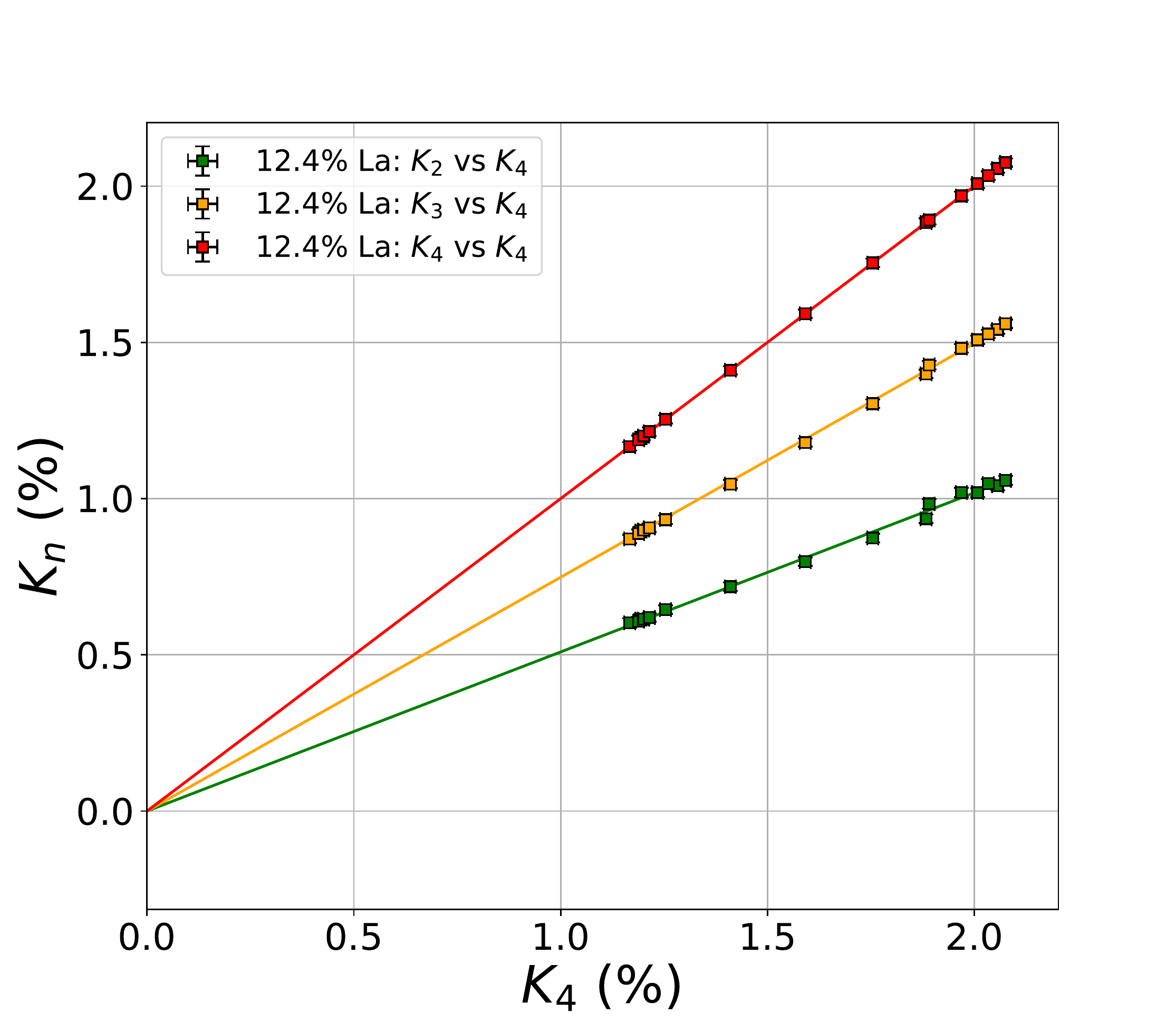}
    \includegraphics[width=\linewidth]{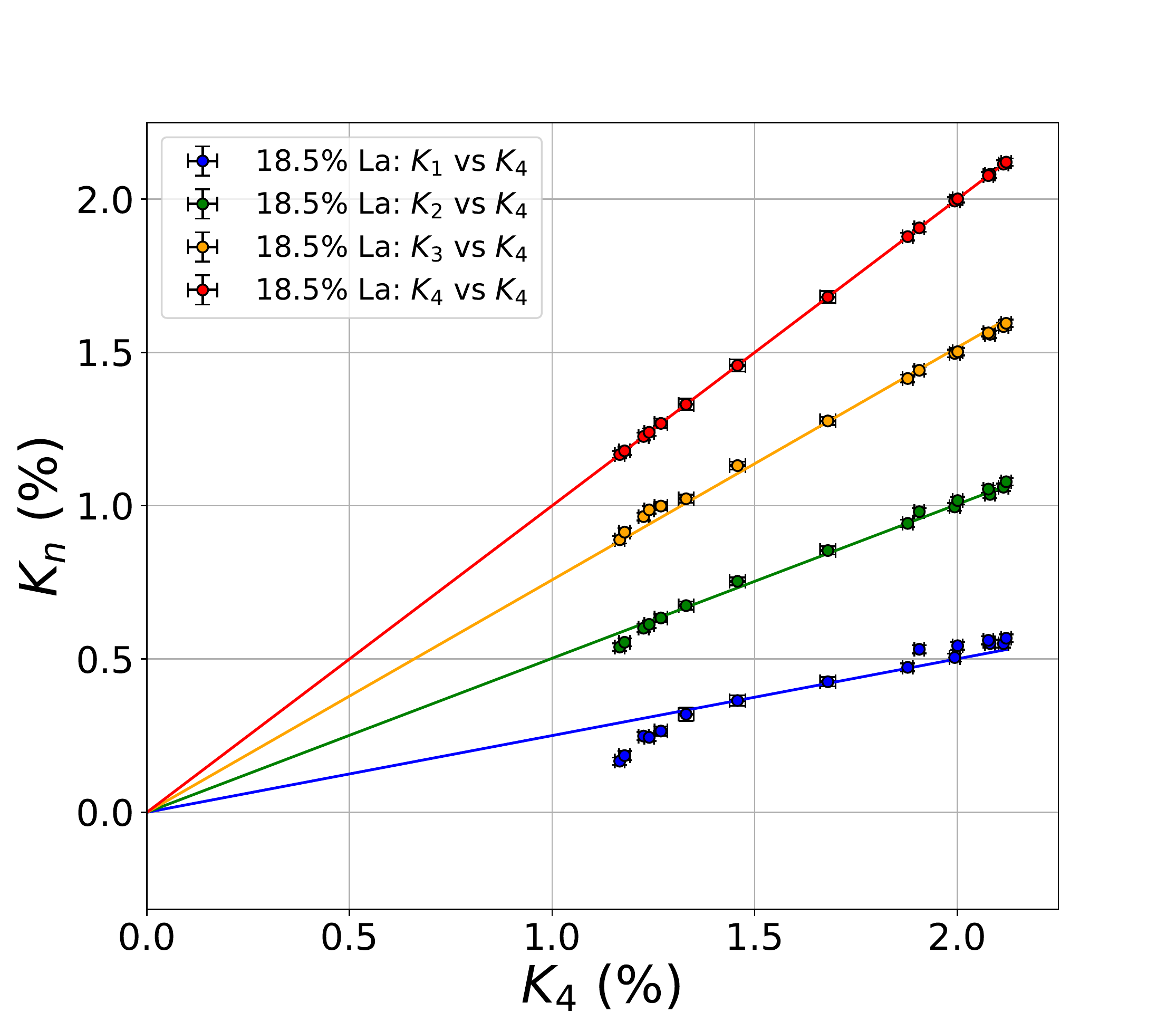}
    \caption{Knight shifts $K_n$ versus $K_4$, for both $x=11.9$\% (upper) and for $x=18.4$\% (lower). The solid lines are fits as described in the text.}
    \label{fig:KvsK}
\end{figure}

Because there are two types of electron spins, $\mathbf{S}_c$ and $\mathbf{S}_f$, there are three distinct components of magnetic susceptibilities, $\chi_{cc}$, $\chi_{cf}$ and $\chi_{ff}$.   The hyperfine interaction is given by: $\mathcal{H}_{\rm hf} = \mathbf{\hat{I}}\cdot[A \mathbf{S}_c + B\sum_{i\in n.n.} \mathbf{S}_f(\mathbf{r}_i)$], where $A$ and $B$ are the hyperfine couplings to the itinerant conduction electron spin and the local $f$-moment, respectively, and the sum is over the four nearest neighbor Ce moments to the central In(1) site \cite{Curro2009}. The Knight shift is given by:
\begin{equation}
    K_n =  A \chi_{cc} + n(A + B) \chi_{cf} + n B \chi_{ff} + K_{0,n}
    \label{eq:KS}
\end{equation}
where $K_{0,n}$ is the temperature independent terms arising from diamagnetic
and orbital contributions \cite{Curro2004,Curro2009,ShirerPNAS2012}, and the bulk susceptibility is given by:
\begin{equation}
\chi = \chi_{cc} + 2 (1 - x) \chi_{cf} + (1 - x) \chi_{ff},
\end{equation}
where $1-x$ is the fraction of Ce spins in the diluted sample.  The different peaks observed in Fig. \ref{fig:x124_spect} can thus be identified by the different $n$, enabling us to  spectrographically distinguish the various types of impurity sites possible in a randomly doped system (see Fig. \ref{fig:unitcell}).

For sufficiently high temperatures, where the correlations between the local moments and the conduction electron spins are negligible, we expect $\chi_{ff}\gg\chi_{cc},\chi_{cf}$ \cite{Curro2004,jiang14}.  In this case  $K_n = K_{0,n} + n B \chi/(1-x)$.  Fig. \ref{fig:KvsT}(b) shows $K_n$ versus $\chi/(1-x)$, which reveals linear behavior for $T\gtrsim 60K$ ($\chi \lesssim 0.008$ emu/mol Ce).  The solid lines show the best linear fits to this data, with the constraint that $B$ is the same for all data sets $K_n$ for a particular crystal.  The fitted values of  $K_{0,n}$ and $B$ are summarized in Table \ref{tab:couplings}. The origin of the  constant term $K_{0,n}$ is not well understood, but it is curious that these values are approximately linearly dependent on $n$. $B$ displays a variability of approximately 9\% between samples, and is consistent with previously reported values in pure CeCoIn$_5$.   In the antiferromagnetic isostructural analog compound CeRhIn$_5$, $B$ is strongly pressure dependent, decreasing by a factor of 3.4 between ambient pressure and 2.0 GPa \cite{Lin2015}.  These results were interpreted as arising from changes in the hybridization as CeRhIn$_5$ is tuned through a quantum critical point.  By 2.0 GPa, antiferromagnetic order has been suppressed in CeRhIn$_5$ and superconductivity emerges, so that this material behaves similarly to CeCoIn$_5$ electronically \cite{tuson}. Our observations in La-doped CeCoIn$_5$ suggest locally-induced strains around the La dopants do not significantly alter the hybridization between the Ce 4f and In 5p orbitals.

\subsection{Fits to Two-Fluid Model}

With the knowledge of $K_{0,n}$ and $B$ determined from the high temperature fits we can now decompose the contributions of the different susceptibilities, $\chi_{\alpha\beta}$.  We define:
\begin{eqnarray}
    \Delta K_n &=& K_n - \frac{n B}{1 - x} \chi - K_{0,n} \\
    &=& n(A - B) \chi_{cf} + \left(A - \frac{n B}{1 - x}\right) \chi_{cc}.
    \label{eq:DeltaK}
\end{eqnarray}
This quantity depends only on $\chi_{cf}$ and $\chi_{cc}$, and is shown in Fig. \ref{fig:deltaK}. The $\Delta K_n$ grow in magnitude at lower temperature, reflecting the growth of correlations between the ${S}_c$ and ${S}_f$ spins at each of the $n$ sites.

The two-fluid model of the Kondo lattice offers a phenomenological framework to describe the behavior of the susceptibility and Knight shift in terms of a set of local $f$-moments and a sea of hybridized heavy electrons \cite{NPF,YangPinesNature,YangPinesPNAS2012,YangDavidPRL}.  This model postulates that  $\Delta K(T)$ is proportional to the susceptibility of the heavy electron fluid, $\chi_{HF}$, and it's temperature dependence probes both growth of hybridization and its relative spectral weight.
We thus fit $\Delta K_n$ to the Yang-Pines expression:
\begin{equation}
\Delta K_n(T) = \Delta K_n^0 \left(1-\frac{T}{T_n^*}\right)^{3/2}\left(1 + \log\frac{T_n^*}{T}\right)
\label{eqn:YangPines}
\end{equation}
to determine the doping ($x$) and site ($n$) dependence of $T_n^*$, as displayed in Fig. \ref{fig:Tstar}.    In fact, $T^*$ is suppressed with dilution, $x$, in approximately the same fashion as observed previously via bulk measurements \cite{NakatsujiFisk}, reflecting a suppression of coherence as intersite couplings are systematically reduced in the dilute lattice.  These values are consistent with previous measurements of $T^*$ in pure CeCoIn$_5$ \cite{Curro2004}, yet are consistently about 30\% higher than reported previously as measured by specific heat and bulk susceptibility.  This discrepancy is likely due to differences in measurement techniques. For a given dilution, $T_n^*$  appears to decrease for the most dilute sites ($n=1$) reflecting local electronic inhomogeneity, as seen in the inset of Fig. \ref{fig:Tstar} \cite{Bauer2011,ParkDropletsNature2013}.  The correlation functions $\chi_{\alpha\beta}$ are expected to become position dependent because translation symmetry is broken in the diluted lattice.  Therefore it is not surprising that different behavior is observed at the different $n$ sites.

\begin{figure}
\centering
    \includegraphics[width=\linewidth]{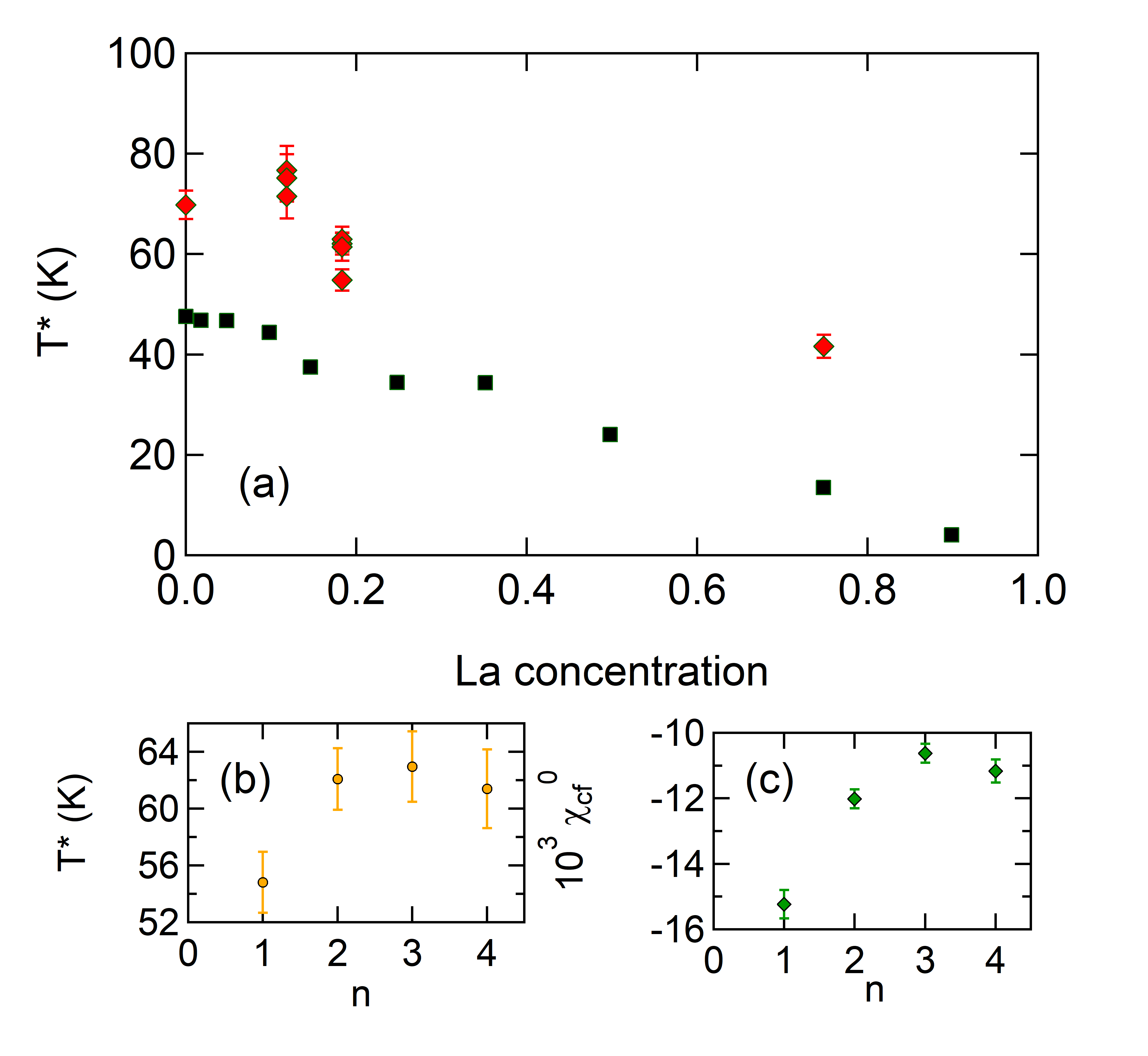}
    \caption{(a) The fitted values of $T_n^*$  versus La concentration, $x$, using Eq. \ref{eqn:YangPines}. The solid squares ($\blacksquare$) are reproduced from \cite{NakatsujiFisk}. (b) and (c) display $T_n^*$  and $\chi_{cf}^0$, respectively, versus site index, $n$, for the $x=$18.5\% sample.}
    \label{fig:Tstar}
\end{figure}

Further evidence for electronic inhomogeneity is observed in the dramatic difference between the $n=0$ and $n=1$ sites for the $x=75\%$ sample \cite{Bauer2011}. $\Delta K_0(T) \approx 0$ the $n=0$ site for the $x=75\%$ sample, whereas for the $n=1$ site the behavior is nearly identical to the bulk. Apparently the heavy electron fluid is not uniformly diluted, but rather becomes spatially varying such that it remains nearly identical to that of the  bulk CeCoIn$_5$ in regions close to the $f$-sites, but vanishes in the intervening regions surrounded by the La. If the length scale of the Kondo screening extended well past the Ce clusters, as theoretical studies would suggest \cite{AffleckKondoScreening}, then the Knight shift of the $n=0$ sites would develop some temperature dependence below $T^*$ in contrast to our observations. Furthermore, the length scale of this inhomogeneity must be relatively short in order to survive such high levels of dilution.  A possible route to understanding the origin of this inhomogeneity may lie in the extent of the Kondo screening clouds surrounding each $f$-site. Theoretical studies indicate that the coherence temperature can increase because the  screening clouds of individual sites overlap forming inter-impurity spin singlets \cite{OgataDilutionPRB2010,ZhuCoupledImpurities2011}.  In some cases $T^*$ can be enhanced by up to an order of magnitude and scale as the RKKY coupling \cite{YangPinesNature}. It remains unclear how many coupled sites are necessary to enhance $T^*$ and what role the dimensionality or network topology of couplings play, however.  For a simple cubic lattice the site percolation limit in 3D is $x_c = 0.31$;  for a 2D square lattice the limit is $x_c = 0.593$ \cite{Percolation3D,*Percolation2Dpart2}.  The Ce lattice in our \celacoin\ sample lies well below the percolation limit; therefore  the occupied Ce sites form disconnected  filamentary clusters of varying sizes, with an average cluster size of 16.4 sites in 3D (1.9 sites per cluster in 2D).

\subsection{Extracting Individual Components}

\begin{figure}
\centering
    \includegraphics[width=\linewidth]{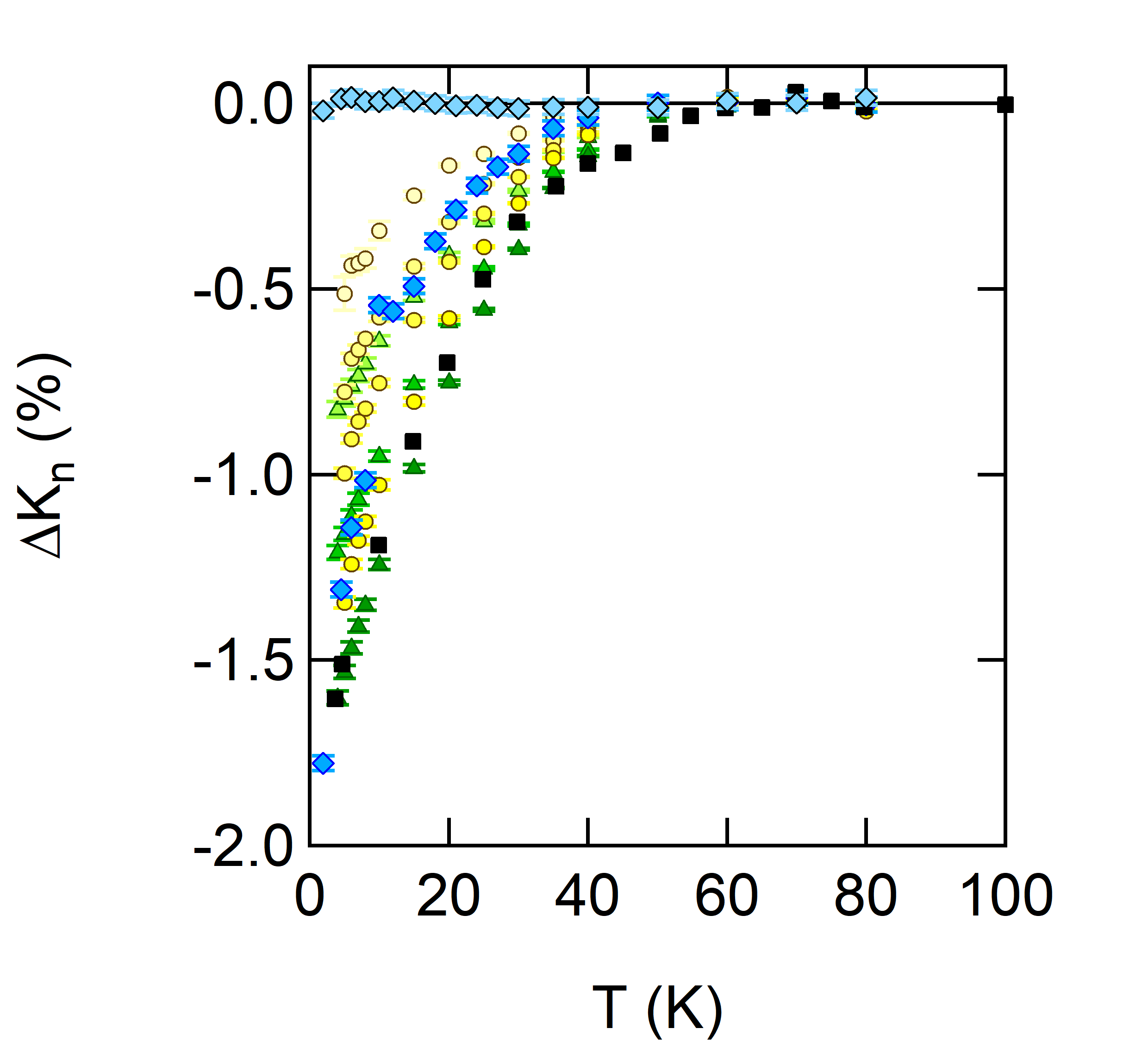}
    \caption{$\Delta K_n$ versus temperature for all La dilutions, $x$, where $\Delta K_n$ is defined in Eq. \ref{eq:DeltaK}. The symbols are identical to those defined in Fig. \ref{fig:KvsT}.}
    \label{fig:deltaK}
\end{figure}

A more complete interpretation of the NMR data has been hampered by the fact that the hyperfine coupling, $A$, to the $S_c$ spins is almost impossible to extract from just Knight shift and susceptibility data.  However, the $\Delta K_n(T)$ data presented in Fig. \ref{fig:deltaK} reveal in interesting trend.   It is apparent that the different $\Delta K_n$ approximately scale with one another, which suggests that $\chi_{cc}(T)$ and $\chi_{cf}(T)$ in Eq. \ref{eq:DeltaK} have a similar temperature dependence.
In this case the ratio $\Delta K_n(T)/\Delta K_m(T)$ is temperature-independent and given by:
\begin{equation}
    \frac{\Delta K_n}{\Delta K_m} = \frac{n(A-B) + AR - nBR/(1-x) }{m(A-B) + AR - mBR/(1-x) },
    \label{eqn:deltaKratio}
\end{equation}
where $R = \chi_{cc}(T)/\chi_{cf}(T)$ is assumed to be temperature independent.   Fig. \ref{fig:deltaK_ratios} shows several plots of $\Delta K_n$ versus $\Delta K_m$ for $x = 18\%$, which clearly reveal a linear relationship. This behavior would not be possible if $\chi_{cc}(T)$ and $\chi_{cf}(T)$ had vastly different temperature dependences.  We perform a global $\chi^2$ minimization over all such data sets for a given $x$ to extract values for $A$ and $R$, shown in the inset and reported in Table \ref{tab:couplings}.  Although the error bars for $A$ are larger than those for $B$, the values are consistent within errors between the two La concentrations. However, these values for $A$ are approximately 25 times smaller than those reported in a previous study in the parent compound, with the opposite sign \cite{Curro2004,NPF,NakatsujiFisk}.   It is likely that the difference arises due to different approaches: in the previous work, $A$ was estimated from an analysis of the bulk susceptibility to extract the heavy electron component, whereas in the current approach we utilize a combination of different Knight shifts and bulk susceptibilities.  The fact that $A$ is negative probably indicates a core polarization mechanism, in which the core orbitals of the In(1) become spin polarized due to coupling to hybridized 5p orbitals \cite{mila89,abragambook}.

\begin{figure}
    \centering
    \includegraphics[width=\linewidth]{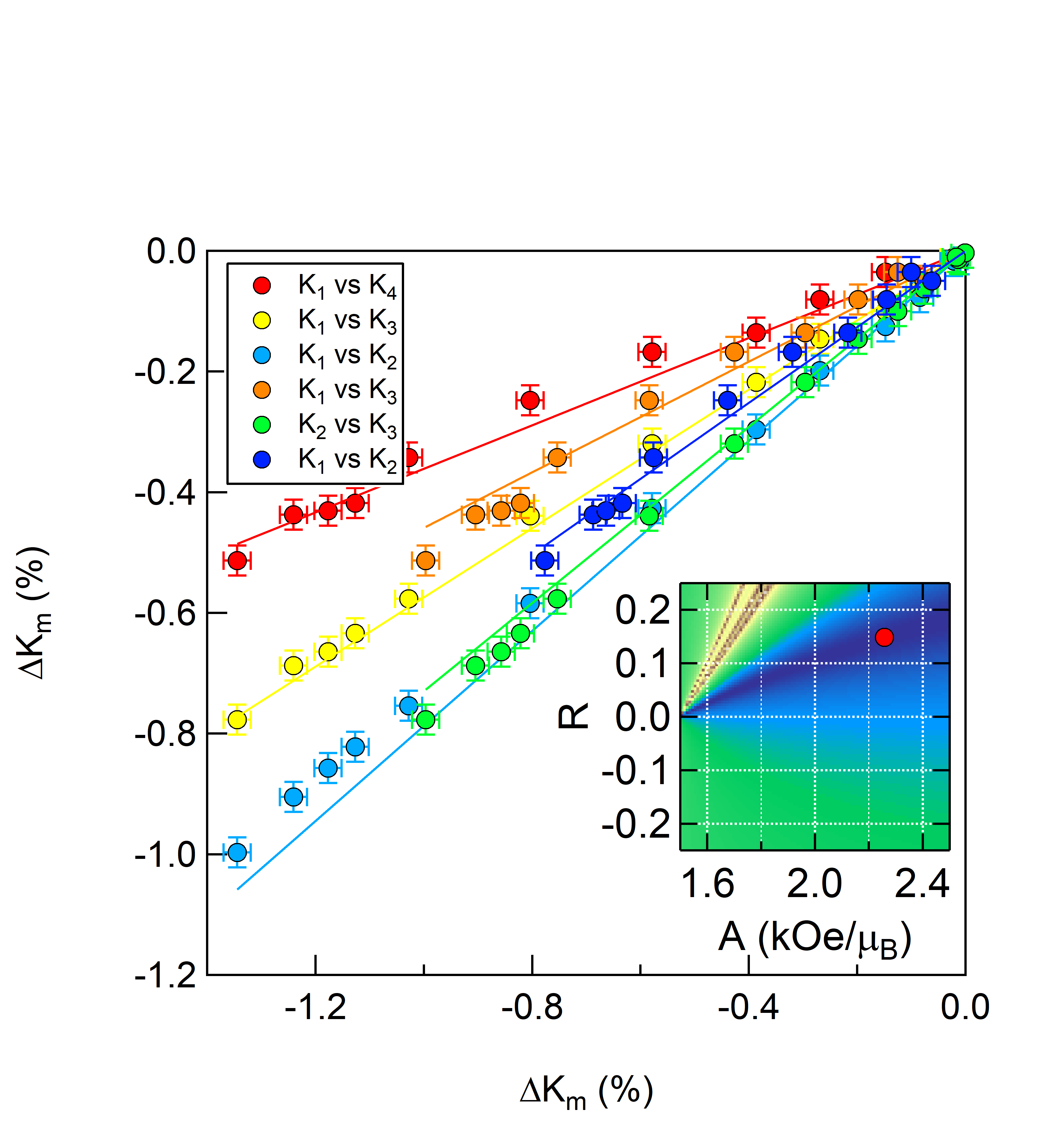}
    \caption{$\Delta K_n$ versus $\Delta K_m$ for all possible combinations of $n$ and $m$ for $x=18\%$. The solid lines are best global fits to the data using Eq. \ref{eqn:deltaKratio}. The inset shows $\chi^2(A,R)$, and the red dot indicates the location of the minimum.}
    \label{fig:deltaK_ratios}
\end{figure}

The specific relationship between the different $\Delta K_n$ in Eq. \ref{eq:DeltaK} provides a unique opportunity to extract the $\chi_{cc}$, $\chi_{cf}$ and $\chi_{ff}$ components independently. Using the values for both hyperfine coupling constants $A$ and $B$ and for the ratio $R$, we can decompose the total susceptibility into individual components as shown in Fig. \ref{fig:linearcombos}. This is the first time these quantities have been measured experimentally - previous studies of the Knight shift anomaly  focused solely on $\Delta K(T)$, but did not determine the on-site coupling, $A$, nor extract the separate contributions of the three components. Note that the large error bars for $R$ may lead to an overestimation of the magnitude of $\chi_{cf}$ and $\chi_{cc}$, however the general behavior of the different susceptibilities agrees qualitatively with theoretical expectations \cite{ShirerPNAS2012,jiang14}.  Namely, $\chi_{cf} < 0$, reflecting the antiferromagnetic nature of the Kondo coupling, whereas $\chi_{cc,ff} > 0$.  For the $x=75.4\%$ sample, only two sites are distinguished: $n=1$ and $n=0$.  Since the $n=0$ site has no temperature dependence, we do not have independent information for either $A$ or $R$.  Similarly, for the $x=0$ data, only the $n=4$ site is present. In these cases we have used the average values of $A$ and $R$ from the $x=12.4$ and $18.5$\% samples to extract the susceptibility components  shown in Fig. \ref{fig:linearcombos} for $x=0$ and $x=75\%$.

\begin{table*}
 \caption{\label{tab:couplings} {The fitted values for the hyperfine couplings $A$ and $B$, and temperature-independent components  $K_{0,n}$ (defined in Eq. \ref{eq:KS}) as well as the ratio $R$ (defined in Eq. \ref{eqn:deltaKratio}) for each of the La dilutions studied.}}
\begin{ruledtabular}
\begin{tabular}{llllllll}
        $x$ & $A$ (kOe/$\mu_B$) & $B$ (kOe/$\mu_B$) & $K_{0, n=1}$ (\%)&  $K_{0, n=2}$ (\%)&  $K_{0, n=3}$  (\%)&  $K_{0, n=4}$ (\%) & $R$\\
        \hline
        $0$ & & 1.527 & & & & 1.029 & \\
        $0.119$ &  $1.79 \pm 0.20$ & $ 1.585 \pm 0.002$ &  & $0.535 \pm 0.001$ &  $0.772 \pm 0.001$ &  $1.008 \pm 0.001$ & $0.02 \pm$ 0.05\\
        $0.184$ & $2.25 \pm 0.20$ & $ 1.495 \pm 0.002$ & $0.313 \pm 0.001$ &  $0.573 \pm 0.001$ &  $0.833 \pm 0.001$ & $1.093 \pm 0.001$ & $0.15 \pm$ 0.05\\
        $0.749$ & & $1.79 \pm 0.42$ &  $0.58 \pm 0.06$  & & & &
         \end{tabular}
\end{ruledtabular}
\end{table*}

\begin{figure*}
\centering
    \includegraphics[width=\linewidth]{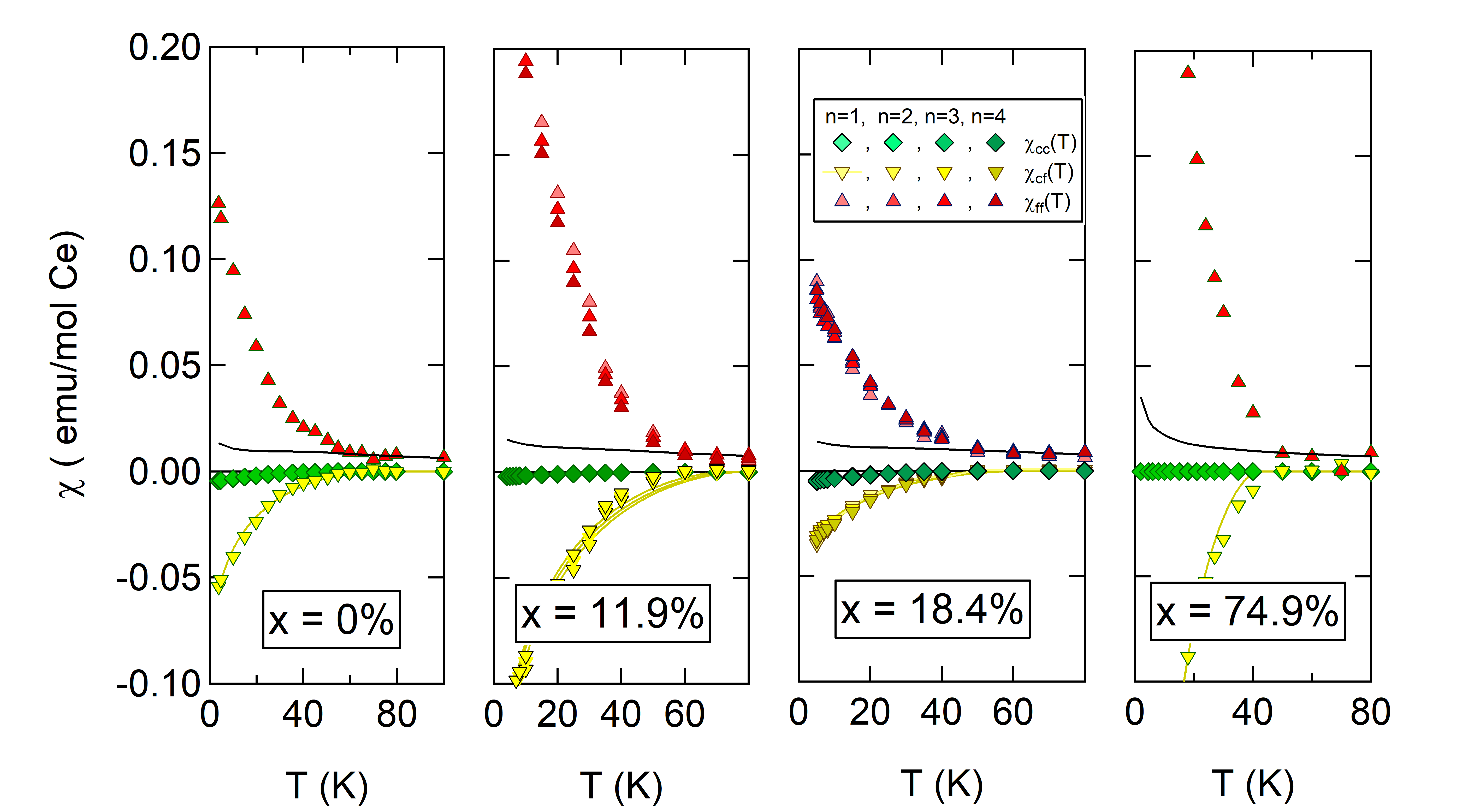}
    \caption{Susceptibility components $\chi_{cc}$ (green, $\diamond$), $\chi_{cf}$ (yellow, $\blacktriangledown$) and $\chi_{ff}$  (red, $\blacktriangle$) for $x=$0, 11.9\%  18.4\% and 74.9\%.  The solid black lines show the bulk susceptibility, $\chi$, and the dashed lines show fits as described in the text. For $x=0$ and 74.9\%, values for $A$ and $R$ were taken as averages of those determined for the other two concentrations.}
    \label{fig:linearcombos}
\end{figure*}

Given these values for $A$ and $B$, we are able to extract the magnitude of $\chi_{cf}^0$, which is shown in Fig. \ref{fig:Tstar}(c).Curiously, the magnitude of $\chi_{cf}^0$ increases at more dilute sites, whereas $T^*$ decreases. The reason for this behavior is unclear, but a naive interpretation is that locally the system is tuned away from quantum criticality, so that the correlation functions are slightly altered \cite{jiang14,ParkDropletsNature2013,BenAliPAMdopants2016}.  On the other hand, $T_1^{-1}$ results suggest otherwise, as discussed below.

\section{Spin Lattice Relaxation Measurements}

\begin{figure}
\centering
    \includegraphics[width=\linewidth]{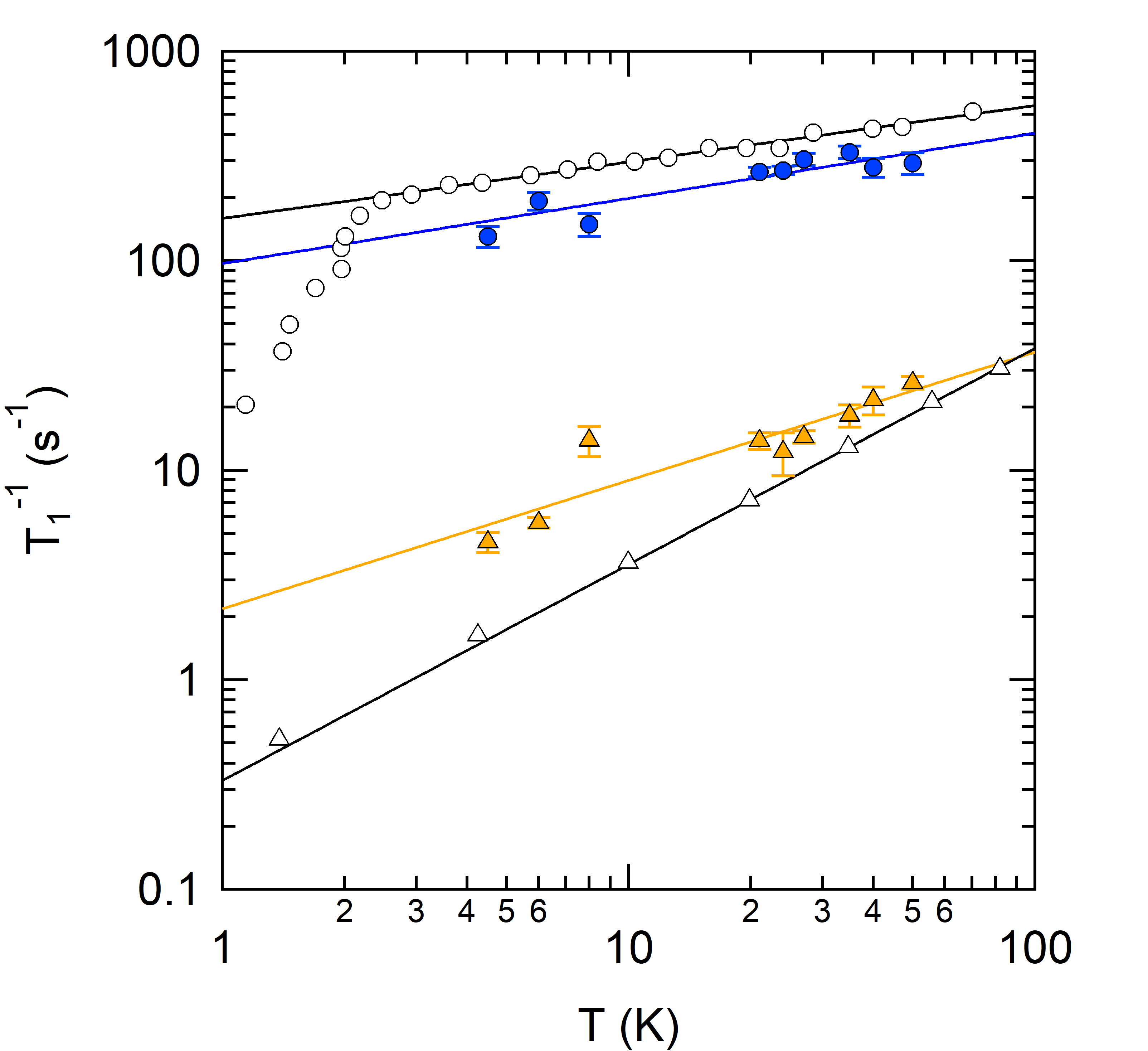}
    \caption{\slrr\ versus temperature for the $n=1$ site (blue, $\bullet$) and $n=0$ site (orange, $\blacktriangle$) for $x=75\%$, compared with the pure CeCoIn$_5$ ($\circ$) and pure LaCoIn$_5$ ($\vartriangle$).  Data for the pure compounds is reproduced from \cite{kohori115shift}.}
    \label{fig:tone}
\end{figure}

Fig. \ref{fig:tone} presents \slrr\ measured for both the $n=1$ and $n=0$ sites in the $x=75\%$ crystal.  \slrr\ is significantly larger for the $n=1$ site, as might be expected given that $K_1(T) > K_0(T)$ for this material.  \slrr\ for the $n=1$ site exhibits a temperature dependence that is similar to that of pure CeCoIn$_5$, albeit with a reduced magnitude that likely reflects the reduced hyperfine coupling.  \slrr\ for the $n=0$ site is approximately one order of magnitude smaller than for the $n=1$ site, and is similar to that observed for the pure LaCoIn$_5$.

Several studies of pure CeCoIn$_5$ indicate this material's proximity to a quantum critical point \cite{romanQCPCeCoIn5,Urbano2007,Yang2009a}. In the NMR response, these fluctuations are manifest in the temperature dependence of \slrr$\sim T^{\alpha}$, where $\alpha = 1/4$ \cite{kohori115shift}. As shown in Fig. \ref{fig:tone}, we find that $\alpha=0.27\pm0.02$ and $0.31\pm0.06$ for the pure CeCoIn$_5$ and the clusters in \celacoin, respectively. The fact that these critical spin fluctuations remain present in short filamentary clusters down to the scale of a few lattice sites suggests that either the coherence length of these fluctuations  is less than the cluster size or that the critical spin fluctuations are local in nature \cite{SiLocalQCP}.  Similar conclusions have been drawn from neutron scattering experiments in CeCu$_{1-x}$Au$_x$ in which the critical fluctuations are independent of wavevector \cite{Schroder2000}. For the $n=0$ site, we find $\alpha = 0.61\pm 0.12$, whereas pure LaCoIn$_5$ exhibits $\alpha = 1.03\pm 0.01$, consistent with Korringa behavior for a Fermi liquid. The fluctuations at the  $n=0$ site for $x=75\%$ are not Korringa, suggesting that heavy electron component of the proximal Ce clusters may still affect the dynamics even though there is no evidence in the Knight shift, $K_0$.

\section{Conclusions}

Our NMR studies of \celacoinx\ have uncovered a series of distinct In(1) sites associated with different numbers of nearest-neighbor $f$-sites.  By comparing the Knight shifts of these sites with the bulk susceptibility and with one another, we extract the temperature dependence of the three correlation functions, $\chi_{cc}$, $\chi_{cf}$, and $\chi_{ff}$ independently.  The susceptibility of the heavy electron fluid, which is a linear combination of $\chi_{cc}$ and $\chi_{cf}$, is systematically reduced with La dilution, as is the coherence temperature, $T^*$.  However, the heavy electron component also becomes spatially inhomogeneous and vanishes in the local vicinity of the La, with a length scale that is on the order of a lattice constant. These results are consistent with recent determinant quantum Monte Carlo (DQMC) calculations of the periodic Anderson model, albeit at half filling, which found that the doping-induced changes in the electronic state are  limited to the nearest-neighbor sites and decay rapidly at the next-nearest-neighbor site \cite{Wei2017}. The \slrrtext\ is inhomogeneous, reflecting quantum critical fluctuations for sites coupled to nearest neighbor $f$-sites, but little or no such fluctuations for sites with no $f$ neighbors.  The suppression of $T^*$ with dilution indicates that intersite couplings among the $f$-spins are important for the emergence of heavy electron coherence in clusters, but the local spin correlations are suppressed at the boundaries of these clusters. Future studies of dilution in related materials such as CeRhIn$_5$ may shed light on how this inhomogeneity evolves away from the quantum critical point.

\acknowledgments{We gratefully  thank C. Callaway, N. Costa, P. Klavins, A. Lodhia, D. Pines, T. Santos, R. Scalettar,  Y. Yang and {J.-X. Zhu} for stimulating discussions,  N. Botto and Dr. S. Roeske for assistance with the microprobe analysis, and D. Hemer for aid with the probe design. Work at UC Davis was supported by the the NSF under Grant No.\ DMR-1005393 and DMR-1807889.}

\bibliography{CeLaCoIn5PRBreferences}

\begin{thebibliography}{44}%
\makeatletter
\providecommand \@ifxundefined [1]{%
 \@ifx{#1\undefined}
}%
\providecommand \@ifnum [1]{%
 \ifnum #1\expandafter \@firstoftwo
 \else \expandafter \@secondoftwo
 \fi
}%
\providecommand \@ifx [1]{%
 \ifx #1\expandafter \@firstoftwo
 \else \expandafter \@secondoftwo
 \fi
}%
\providecommand \natexlab [1]{#1}%
\providecommand \enquote  [1]{``#1''}%
\providecommand \bibnamefont  [1]{#1}%
\providecommand \bibfnamefont [1]{#1}%
\providecommand \citenamefont [1]{#1}%
\providecommand \href@noop [0]{\@secondoftwo}%
\providecommand \href [0]{\begingroup \@sanitize@url \@href}%
\providecommand \@href[1]{\@@startlink{#1}\@@href}%
\providecommand \@@href[1]{\endgroup#1\@@endlink}%
\providecommand \@sanitize@url [0]{\catcode `\\12\catcode `\$12\catcode
  `\&12\catcode `\#12\catcode `\^12\catcode `\_12\catcode `\%12\relax}%
\providecommand \@@startlink[1]{}%
\providecommand \@@endlink[0]{}%
\providecommand \url  [0]{\begingroup\@sanitize@url \@url }%
\providecommand \@url [1]{\endgroup\@href {#1}{\urlprefix }}%
\providecommand \urlprefix  [0]{URL }%
\providecommand \Eprint [0]{\href }%
\providecommand \doibase [0]{http://dx.doi.org/}%
\providecommand \selectlanguage [0]{\@gobble}%
\providecommand \bibinfo  [0]{\@secondoftwo}%
\providecommand \bibfield  [0]{\@secondoftwo}%
\providecommand \translation [1]{[#1]}%
\providecommand \BibitemOpen [0]{}%
\providecommand \bibitemStop [0]{}%
\providecommand \bibitemNoStop [0]{.\EOS\space}%
\providecommand \EOS [0]{\spacefactor3000\relax}%
\providecommand \BibitemShut  [1]{\csname bibitem#1\endcsname}%
\let\auto@bib@innerbib\@empty
\bibitem [{\citenamefont {Stewart}(2001)}]{StewartHFreview}%
  \BibitemOpen
  \bibfield  {author} {\bibinfo {author} {\bibfnamefont {G.~R.}\ \bibnamefont
  {Stewart}},\ }\bibfield  {title} {\enquote {\bibinfo {title}
  {Non-{Fermi}-liquid behavior in $d$- and $f$-electron metals},}\ }\href
  {\doibase 10.1103/RevModPhys.73.797} {\bibfield  {journal} {\bibinfo
  {journal} {Rev. Mod. Phys.}\ }\textbf {\bibinfo {volume} {73}},\ \bibinfo
  {pages} {797--855} (\bibinfo {year} {2001})}\BibitemShut {NoStop}%
\bibitem [{\citenamefont {Coleman}\ and\ \citenamefont
  {Schofield}(2005)}]{ColemanQCreview}%
  \BibitemOpen
  \bibfield  {author} {\bibinfo {author} {\bibfnamefont {Piers}\ \bibnamefont
  {Coleman}}\ and\ \bibinfo {author} {\bibfnamefont {Andrew~J.}\ \bibnamefont
  {Schofield}},\ }\bibfield  {title} {\enquote {\bibinfo {title} {Quantum
  criticality},}\ }\href {http://dx.doi.org/10.1038/nature03279} {\bibfield
  {journal} {\bibinfo  {journal} {Nature}\ }\textbf {\bibinfo {volume} {433}},\
  \bibinfo {pages} {226--229} (\bibinfo {year} {2005})}\BibitemShut {NoStop}%
\bibitem [{\citenamefont {Hewson}(1993)}]{hewson}%
  \BibitemOpen
  \bibfield  {author} {\bibinfo {author} {\bibfnamefont {A.~C.}\ \bibnamefont
  {Hewson}},\ }\href@noop {} {\emph {\bibinfo {title} {The Kondo Problem to
  Heavy Fermions}}}\ (\bibinfo  {publisher} {Cambridge University Press},\
  \bibinfo {year} {1993})\BibitemShut {NoStop}%
\bibitem [{\citenamefont {Jones}\ and\ \citenamefont
  {Varma}(1987)}]{JonesVarmaPRL1987}%
  \BibitemOpen
  \bibfield  {author} {\bibinfo {author} {\bibfnamefont {B.~A.}\ \bibnamefont
  {Jones}}\ and\ \bibinfo {author} {\bibfnamefont {C.~M.}\ \bibnamefont
  {Varma}},\ }\bibfield  {title} {\enquote {\bibinfo {title} {Study of two
  magnetic impurities in a {F}ermi gas},}\ }\href {\doibase
  10.1103/PhysRevLett.58.843} {\bibfield  {journal} {\bibinfo  {journal} {Phys.
  Rev. Lett.}\ }\textbf {\bibinfo {volume} {58}},\ \bibinfo {pages} {843--846}
  (\bibinfo {year} {1987})}\BibitemShut {NoStop}%
\bibitem [{\citenamefont {Jones}\ \emph {et~al.}(1988)\citenamefont {Jones},
  \citenamefont {Varma},\ and\ \citenamefont
  {Wilkins}}]{VarmaJonesTwoKondoPRL}%
  \BibitemOpen
  \bibfield  {author} {\bibinfo {author} {\bibfnamefont {B.~A.}\ \bibnamefont
  {Jones}}, \bibinfo {author} {\bibfnamefont {C.~M.}\ \bibnamefont {Varma}}, \
  and\ \bibinfo {author} {\bibfnamefont {J.~W.}\ \bibnamefont {Wilkins}},\
  }\bibfield  {title} {\enquote {\bibinfo {title} {Low-temperature properties
  of the two-impurity {K}ondo {H}amiltonian},}\ }\href {\doibase
  10.1103/PhysRevLett.61.125} {\bibfield  {journal} {\bibinfo  {journal} {Phys.
  Rev. Lett.}\ }\textbf {\bibinfo {volume} {61}},\ \bibinfo {pages} {125--128}
  (\bibinfo {year} {1988})}\BibitemShut {NoStop}%
\bibitem [{\citenamefont {Fye}\ \emph {et~al.}(1987)\citenamefont {Fye},
  \citenamefont {Hirsch},\ and\ \citenamefont {Scalapino}}]{HirschTwoKondoPRB}%
  \BibitemOpen
  \bibfield  {author} {\bibinfo {author} {\bibfnamefont {R.~M.}\ \bibnamefont
  {Fye}}, \bibinfo {author} {\bibfnamefont {J.~E.}\ \bibnamefont {Hirsch}}, \
  and\ \bibinfo {author} {\bibfnamefont {D.~J.}\ \bibnamefont {Scalapino}},\
  }\bibfield  {title} {\enquote {\bibinfo {title} {Kondo effect versus indirect
  exchange in the two-impurity {A}nderson model: A {Monte Carlo} study},}\
  }\href {\doibase 10.1103/PhysRevB.35.4901} {\bibfield  {journal} {\bibinfo
  {journal} {Phys. Rev. B}\ }\textbf {\bibinfo {volume} {35}},\ \bibinfo
  {pages} {4901--4908} (\bibinfo {year} {1987})}\BibitemShut {NoStop}%
\bibitem [{\citenamefont {Doniach}(1977)}]{doniach}%
  \BibitemOpen
  \bibfield  {author} {\bibinfo {author} {\bibfnamefont {S.}~\bibnamefont
  {Doniach}},\ }\bibfield  {title} {\enquote {\bibinfo {title} {The {K}ondo
  lattice and weak antiferromagnetism},}\ }\href {\doibase
  10.1016/0378-4363(77)90190-5} {\bibfield  {journal} {\bibinfo  {journal}
  {Physica}\ }\textbf {\bibinfo {volume} {91B}},\ \bibinfo {pages} {231--234}
  (\bibinfo {year} {1977})}\BibitemShut {NoStop}%
\bibitem [{\citenamefont {Watanabe}\ and\ \citenamefont
  {Ogata}(2010)}]{OgataDilutionPRB2010}%
  \BibitemOpen
  \bibfield  {author} {\bibinfo {author} {\bibfnamefont {Hiroshi}\ \bibnamefont
  {Watanabe}}\ and\ \bibinfo {author} {\bibfnamefont {Masao}\ \bibnamefont
  {Ogata}},\ }\bibfield  {title} {\enquote {\bibinfo {title} {Crossover from
  dilute {K}ondo system to heavy-fermion system},}\ }\href {\doibase
  10.1103/PhysRevB.81.113111} {\bibfield  {journal} {\bibinfo  {journal} {Phys.
  Rev. B}\ }\textbf {\bibinfo {volume} {81}},\ \bibinfo {pages} {113111}
  (\bibinfo {year} {2010})}\BibitemShut {NoStop}%
\bibitem [{\citenamefont {Zhu}\ and\ \citenamefont
  {Zhu}(2011)}]{ZhuCoupledImpurities2011}%
  \BibitemOpen
  \bibfield  {author} {\bibinfo {author} {\bibfnamefont {Lijun}\ \bibnamefont
  {Zhu}}\ and\ \bibinfo {author} {\bibfnamefont {Jian-Xin}\ \bibnamefont
  {Zhu}},\ }\bibfield  {title} {\enquote {\bibinfo {title} {Coherence scale of
  coupled {A}nderson impurities},}\ }\href {\doibase
  10.1103/PhysRevB.83.195103} {\bibfield  {journal} {\bibinfo  {journal} {Phys.
  Rev. B}\ }\textbf {\bibinfo {volume} {83}},\ \bibinfo {pages} {195103}
  (\bibinfo {year} {2011})}\BibitemShut {NoStop}%
\bibitem [{\citenamefont {Custers}\ \emph {et~al.}(2003)\citenamefont
  {Custers}, \citenamefont {Gegenwart}, \citenamefont {Wilhelm}, \citenamefont
  {Neumaier}, \citenamefont {Tokiwa}, \citenamefont {Trovarelli}, \citenamefont
  {Geibel}, \citenamefont {Steglich}, \citenamefont {Pepin},\ and\
  \citenamefont {Coleman}}]{YRSnature}%
  \BibitemOpen
  \bibfield  {author} {\bibinfo {author} {\bibfnamefont {J.}~\bibnamefont
  {Custers}}, \bibinfo {author} {\bibfnamefont {P.}~\bibnamefont {Gegenwart}},
  \bibinfo {author} {\bibfnamefont {H.}~\bibnamefont {Wilhelm}}, \bibinfo
  {author} {\bibfnamefont {K.}~\bibnamefont {Neumaier}}, \bibinfo {author}
  {\bibfnamefont {Y.}~\bibnamefont {Tokiwa}}, \bibinfo {author} {\bibfnamefont
  {O.}~\bibnamefont {Trovarelli}}, \bibinfo {author} {\bibfnamefont
  {C.}~\bibnamefont {Geibel}}, \bibinfo {author} {\bibfnamefont
  {F.}~\bibnamefont {Steglich}}, \bibinfo {author} {\bibfnamefont
  {C.}~\bibnamefont {Pepin}}, \ and\ \bibinfo {author} {\bibfnamefont
  {P.}~\bibnamefont {Coleman}},\ }\bibfield  {title} {\enquote {\bibinfo
  {title} {The break-up of heavy electrons at a quantum critical point},}\
  }\href {\doibase 10.1038/nature01774} {\bibfield  {journal} {\bibinfo
  {journal} {Nature}\ }\textbf {\bibinfo {volume} {424}},\ \bibinfo {pages}
  {524--527} (\bibinfo {year} {2003})}\BibitemShut {NoStop}%
\bibitem [{\citenamefont {Coleman}\ \emph {et~al.}(2001)\citenamefont
  {Coleman}, \citenamefont {Pepin}, \citenamefont {Si},\ and\ \citenamefont
  {Ramazashvili}}]{ColemanHFdeath}%
  \BibitemOpen
  \bibfield  {author} {\bibinfo {author} {\bibfnamefont {P.}~\bibnamefont
  {Coleman}}, \bibinfo {author} {\bibfnamefont {C.}~\bibnamefont {Pepin}},
  \bibinfo {author} {\bibfnamefont {Q.~M.}\ \bibnamefont {Si}}, \ and\ \bibinfo
  {author} {\bibfnamefont {R.}~\bibnamefont {Ramazashvili}},\ }\bibfield
  {title} {\enquote {\bibinfo {title} {How do {F}ermi liquids get heavy and
  die?.}}\ }\href {\doibase 10.1088/0953-8984/13/35/202} {\bibfield  {journal}
  {\bibinfo  {journal} {J. Phys.: Condens. Matter}\ }\textbf {\bibinfo {volume}
  {13}},\ \bibinfo {pages} {R723 -- 38} (\bibinfo {year} {2001})}\BibitemShut
  {NoStop}%
\bibitem [{\citenamefont {Bork}\ \emph {et~al.}(2011)\citenamefont {Bork},
  \citenamefont {Zhang}, \citenamefont {Diekhoner}, \citenamefont {Borda},
  \citenamefont {Simon}, \citenamefont {Kroha}, \citenamefont {Wahl},\ and\
  \citenamefont {Kern}}]{STM2KondoNature2011}%
  \BibitemOpen
  \bibfield  {author} {\bibinfo {author} {\bibfnamefont {Jakob}\ \bibnamefont
  {Bork}}, \bibinfo {author} {\bibfnamefont {Yong-hui}\ \bibnamefont {Zhang}},
  \bibinfo {author} {\bibfnamefont {Lars}\ \bibnamefont {Diekhoner}}, \bibinfo
  {author} {\bibfnamefont {Laszlo}\ \bibnamefont {Borda}}, \bibinfo {author}
  {\bibfnamefont {Pascal}\ \bibnamefont {Simon}}, \bibinfo {author}
  {\bibfnamefont {Johann}\ \bibnamefont {Kroha}}, \bibinfo {author}
  {\bibfnamefont {Peter}\ \bibnamefont {Wahl}}, \ and\ \bibinfo {author}
  {\bibfnamefont {Klaus}\ \bibnamefont {Kern}},\ }\bibfield  {title} {\enquote
  {\bibinfo {title} {A tunable two-impurity {K}ondo system in an atomic point
  contact},}\ }\href {http://dx.doi.org/10.1038/nphys2076} {\bibfield
  {journal} {\bibinfo  {journal} {Nat. Phys.}\ }\textbf {\bibinfo {volume}
  {7}},\ \bibinfo {pages} {901--906} (\bibinfo {year} {2011})}\BibitemShut
  {NoStop}%
\bibitem [{\citenamefont {Petrovic}\ \emph {et~al.}(2001)\citenamefont
  {Petrovic}, \citenamefont {Pagliuso}, \citenamefont {Hundley}, \citenamefont
  {Movshovich}, \citenamefont {Sarrao}, \citenamefont {Thompson}, \citenamefont
  {Fisk},\ and\ \citenamefont {Monthoux}}]{CeCoIn5discovery}%
  \BibitemOpen
  \bibfield  {author} {\bibinfo {author} {\bibfnamefont {C.}~\bibnamefont
  {Petrovic}}, \bibinfo {author} {\bibfnamefont {P.~G.}\ \bibnamefont
  {Pagliuso}}, \bibinfo {author} {\bibfnamefont {M.~F.}\ \bibnamefont
  {Hundley}}, \bibinfo {author} {\bibfnamefont {R.}~\bibnamefont {Movshovich}},
  \bibinfo {author} {\bibfnamefont {J.~L.}\ \bibnamefont {Sarrao}}, \bibinfo
  {author} {\bibfnamefont {J.~D.}\ \bibnamefont {Thompson}}, \bibinfo {author}
  {\bibfnamefont {Z.}~\bibnamefont {Fisk}}, \ and\ \bibinfo {author}
  {\bibfnamefont {P.}~\bibnamefont {Monthoux}},\ }\bibfield  {title} {\enquote
  {\bibinfo {title} {Heavy-fermion superconductivity in {CeCoIn$_5$} at 2.3
  {K}},}\ }\href {\doibase 10.1088/0953-8984/13/17/103} {\bibfield  {journal}
  {\bibinfo  {journal} {J. Phys. Cond. Mat.}\ }\textbf {\bibinfo {volume}
  {13}},\ \bibinfo {pages} {L337--L342} (\bibinfo {year} {2001})}\BibitemShut
  {NoStop}%
\bibitem [{\citenamefont {Bianchi}\ \emph {et~al.}(2003)\citenamefont
  {Bianchi}, \citenamefont {Movshovich}, \citenamefont {Vekhter}, \citenamefont
  {Pagliuso},\ and\ \citenamefont {Sarrao}}]{romanQCPCeCoIn5}%
  \BibitemOpen
  \bibfield  {author} {\bibinfo {author} {\bibfnamefont {A.}~\bibnamefont
  {Bianchi}}, \bibinfo {author} {\bibfnamefont {R.}~\bibnamefont {Movshovich}},
  \bibinfo {author} {\bibfnamefont {I.}~\bibnamefont {Vekhter}}, \bibinfo
  {author} {\bibfnamefont {P.~G.}\ \bibnamefont {Pagliuso}}, \ and\ \bibinfo
  {author} {\bibfnamefont {J.~L.}\ \bibnamefont {Sarrao}},\ }\bibfield  {title}
  {\enquote {\bibinfo {title} {Avoided antiferromagnetic order and quantum
  critical point in {CeCoIn$_5$}},}\ }\href@noop {} {\bibfield  {journal}
  {\bibinfo  {journal} {Phys. Rev. Lett.}\ }\textbf {\bibinfo {volume} {91}},\
  \bibinfo {pages} {257001 -- 4} (\bibinfo {year} {2003})}\BibitemShut
  {NoStop}%
\bibitem [{\citenamefont {Sykes}\ \emph
  {et~al.}(1976{\natexlab{a}})\citenamefont {Sykes}, \citenamefont {Gaunt},\
  and\ \citenamefont {Glen}}]{Percolation3D}%
  \BibitemOpen
  \bibfield  {author} {\bibinfo {author} {\bibfnamefont {M~F}\ \bibnamefont
  {Sykes}}, \bibinfo {author} {\bibfnamefont {D~S}\ \bibnamefont {Gaunt}}, \
  and\ \bibinfo {author} {\bibfnamefont {M}~\bibnamefont {Glen}},\ }\bibfield
  {title} {\enquote {\bibinfo {title} {Percolation processes in three
  dimensions},}\ }\href {http://stacks.iop.org/0305-4470/9/i=10/a=021}
  {\bibfield  {journal} {\bibinfo  {journal} {J. Phys. A: Math. Gen.}\ }\textbf
  {\bibinfo {volume} {9}},\ \bibinfo {pages} {1705} (\bibinfo {year}
  {1976}{\natexlab{a}})}\BibitemShut {NoStop}%
\bibitem [{\citenamefont {Sykes}\ \emph
  {et~al.}(1976{\natexlab{b}})\citenamefont {Sykes}, \citenamefont {Gaunt},\
  and\ \citenamefont {Glen}}]{Percolation2Dpart2}%
  \BibitemOpen
  \bibfield  {author} {\bibinfo {author} {\bibfnamefont {M~F}\ \bibnamefont
  {Sykes}}, \bibinfo {author} {\bibfnamefont {D~S}\ \bibnamefont {Gaunt}}, \
  and\ \bibinfo {author} {\bibfnamefont {M}~\bibnamefont {Glen}},\ }\bibfield
  {title} {\enquote {\bibinfo {title} {Percolation processes in two dimensions.
  {II. C}ritical concentrations and the mean size index},}\ }\href
  {http://stacks.iop.org/0305-4470/9/i=1/a=015} {\bibfield  {journal} {\bibinfo
   {journal} {J. Phys. A: Math. Gen.}\ }\textbf {\bibinfo {volume} {9}},\
  \bibinfo {pages} {97} (\bibinfo {year} {1976}{\natexlab{b}})}\BibitemShut
  {NoStop}%
\bibitem [{\citenamefont {Lin}\ \emph {et~al.}(1987)\citenamefont {Lin},
  \citenamefont {Wallash}, \citenamefont {Crow}, \citenamefont {Mihalisin},\
  and\ \citenamefont {Schlottmann}}]{CePb3DilutionPRL}%
  \BibitemOpen
  \bibfield  {author} {\bibinfo {author} {\bibfnamefont {C.~L.}\ \bibnamefont
  {Lin}}, \bibinfo {author} {\bibfnamefont {A.}~\bibnamefont {Wallash}},
  \bibinfo {author} {\bibfnamefont {J.~E.}\ \bibnamefont {Crow}}, \bibinfo
  {author} {\bibfnamefont {T.}~\bibnamefont {Mihalisin}}, \ and\ \bibinfo
  {author} {\bibfnamefont {P.}~\bibnamefont {Schlottmann}},\ }\bibfield
  {title} {\enquote {\bibinfo {title} {Heavy-fermion behavior and the
  single-ion {K}ondo model},}\ }\href {\doibase 10.1103/PhysRevLett.58.1232}
  {\bibfield  {journal} {\bibinfo  {journal} {Phys. Rev. Lett.}\ }\textbf
  {\bibinfo {volume} {58}},\ \bibinfo {pages} {1232--1235} (\bibinfo {year}
  {1987})}\BibitemShut {NoStop}%
\bibitem [{\citenamefont {Nakatsuji}\ \emph {et~al.}(2002)\citenamefont
  {Nakatsuji}, \citenamefont {Yeo}, \citenamefont {Balicas}, \citenamefont
  {Fisk}, \citenamefont {Schlottmann}, \citenamefont {Pagliuso}, \citenamefont
  {Moreno}, \citenamefont {Sarrao},\ and\ \citenamefont
  {Thompson}}]{NakatsujiFisk}%
  \BibitemOpen
  \bibfield  {author} {\bibinfo {author} {\bibfnamefont {S.}~\bibnamefont
  {Nakatsuji}}, \bibinfo {author} {\bibfnamefont {S.}~\bibnamefont {Yeo}},
  \bibinfo {author} {\bibfnamefont {L.}~\bibnamefont {Balicas}}, \bibinfo
  {author} {\bibfnamefont {Z.}~\bibnamefont {Fisk}}, \bibinfo {author}
  {\bibfnamefont {P.}~\bibnamefont {Schlottmann}}, \bibinfo {author}
  {\bibfnamefont {P.~G.}\ \bibnamefont {Pagliuso}}, \bibinfo {author}
  {\bibfnamefont {N.~O.}\ \bibnamefont {Moreno}}, \bibinfo {author}
  {\bibfnamefont {J.~L.}\ \bibnamefont {Sarrao}}, \ and\ \bibinfo {author}
  {\bibfnamefont {J.~D.}\ \bibnamefont {Thompson}},\ }\bibfield  {title}
  {\enquote {\bibinfo {title} {Intersite coupling effects in a {K}ondo
  lattice},}\ }\href {\doibase 10.1103/PhysRevLett.89.106402} {\bibfield
  {journal} {\bibinfo  {journal} {Phys. Rev. Lett.}\ }\textbf {\bibinfo
  {volume} {89}},\ \bibinfo {pages} {106402} (\bibinfo {year}
  {2002})}\BibitemShut {NoStop}%
\bibitem [{\citenamefont {Yang}\ \emph {et~al.}(2008)\citenamefont {Yang},
  \citenamefont {Fisk}, \citenamefont {Lee}, \citenamefont {Thompson},\ and\
  \citenamefont {Pines}}]{YangPinesNature}%
  \BibitemOpen
  \bibfield  {author} {\bibinfo {author} {\bibfnamefont {Y.-F.}\ \bibnamefont
  {Yang}}, \bibinfo {author} {\bibfnamefont {Zachary}\ \bibnamefont {Fisk}},
  \bibinfo {author} {\bibfnamefont {Han-Oh}\ \bibnamefont {Lee}}, \bibinfo
  {author} {\bibfnamefont {J.~D.}\ \bibnamefont {Thompson}}, \ and\ \bibinfo
  {author} {\bibfnamefont {David}\ \bibnamefont {Pines}},\ }\bibfield  {title}
  {\enquote {\bibinfo {title} {Scaling the {K}ondo lattice},}\ }\href {\doibase
  10.1038/nature07157} {\bibfield  {journal} {\bibinfo  {journal} {Nature}\
  }\textbf {\bibinfo {volume} {454}},\ \bibinfo {pages} {611--613} (\bibinfo
  {year} {2008})}\BibitemShut {NoStop}%
\bibitem [{\citenamefont {Aynajian}\ \emph {et~al.}(2012)\citenamefont
  {Aynajian}, \citenamefont {da~Silva~Neto}, \citenamefont {Gyenis},
  \citenamefont {Baumbach}, \citenamefont {Thompson}, \citenamefont {Fisk},
  \citenamefont {Bauer},\ and\ \citenamefont {Yazdani}}]{Aynajian2012}%
  \BibitemOpen
  \bibfield  {author} {\bibinfo {author} {\bibfnamefont {Pegor}\ \bibnamefont
  {Aynajian}}, \bibinfo {author} {\bibfnamefont {Eduardo~H.}\ \bibnamefont
  {da~Silva~Neto}}, \bibinfo {author} {\bibfnamefont {Andras}\ \bibnamefont
  {Gyenis}}, \bibinfo {author} {\bibfnamefont {Ryan~E.}\ \bibnamefont
  {Baumbach}}, \bibinfo {author} {\bibfnamefont {J.~D.}\ \bibnamefont
  {Thompson}}, \bibinfo {author} {\bibfnamefont {Zachary}\ \bibnamefont
  {Fisk}}, \bibinfo {author} {\bibfnamefont {Eric~D.}\ \bibnamefont {Bauer}}, \
  and\ \bibinfo {author} {\bibfnamefont {Ali}\ \bibnamefont {Yazdani}},\
  }\bibfield  {title} {\enquote {\bibinfo {title} {Visualizing heavy fermions
  emerging in a quantum critical {K}ondo lattice},}\ }\href
  {http://dx.doi.org/10.1038/nature11204} {\bibfield  {journal} {\bibinfo
  {journal} {Nature}\ }\textbf {\bibinfo {volume} {486}},\ \bibinfo {pages}
  {201--206} (\bibinfo {year} {2012})}\BibitemShut {NoStop}%
\bibitem [{\citenamefont {Nakatsuji}\ \emph {et~al.}(2004)\citenamefont
  {Nakatsuji}, \citenamefont {Pines},\ and\ \citenamefont {Fisk}}]{NPF}%
  \BibitemOpen
  \bibfield  {author} {\bibinfo {author} {\bibfnamefont {S.}~\bibnamefont
  {Nakatsuji}}, \bibinfo {author} {\bibfnamefont {D.}~\bibnamefont {Pines}}, \
  and\ \bibinfo {author} {\bibfnamefont {Z.}~\bibnamefont {Fisk}},\ }\bibfield
  {title} {\enquote {\bibinfo {title} {Two fluid description of the {K}ondo
  lattice.}}\ }\href {\doibase 10.1103/PhysRevLett.92.016401} {\bibfield
  {journal} {\bibinfo  {journal} {Phys. Rev. Lett.}\ }\textbf {\bibinfo
  {volume} {92}},\ \bibinfo {pages} {016401 -- 4} (\bibinfo {year}
  {2004})}\BibitemShut {NoStop}%
\bibitem [{\citenamefont {Yang}\ and\ \citenamefont
  {Pines}(2012)}]{YangPinesPNAS2012}%
  \BibitemOpen
  \bibfield  {author} {\bibinfo {author} {\bibfnamefont {Yi-feng}\ \bibnamefont
  {Yang}}\ and\ \bibinfo {author} {\bibfnamefont {David}\ \bibnamefont
  {Pines}},\ }\bibfield  {title} {\enquote {\bibinfo {title} {Emergent states
  in heavy-electron materials},}\ }\href {\doibase 10.1073/pnas.1211186109}
  {\bibfield  {journal} {\bibinfo  {journal} {Proc. Natl. Acad. Sci.}\ }\textbf
  {\bibinfo {volume} {109}},\ \bibinfo {pages} {E3060 -- E3066} (\bibinfo
  {year} {2012})}\BibitemShut {NoStop}%
\bibitem [{\citenamefont {Curro}(2009)}]{Curro2009}%
  \BibitemOpen
  \bibfield  {author} {\bibinfo {author} {\bibfnamefont {N~J}\ \bibnamefont
  {Curro}},\ }\bibfield  {title} {\enquote {\bibinfo {title} {Nuclear magnetic
  resonance in the heavy fermion superconductors},}\ }\href {\doibase
  10.1088/0034-4885/72/2/026502} {\bibfield  {journal} {\bibinfo  {journal}
  {Rep. Prog. Phys.}\ }\textbf {\bibinfo {volume} {72}},\ \bibinfo {pages}
  {026502 (24pp)} (\bibinfo {year} {2009})}\BibitemShut {NoStop}%
\bibitem [{\citenamefont {Shirer}\ \emph {et~al.}(2012)\citenamefont {Shirer},
  \citenamefont {Shockley}, \citenamefont {Dioguardi}, \citenamefont {Crocker},
  \citenamefont {Lin}, \citenamefont {apRoberts Warren}, \citenamefont
  {Nisson}, \citenamefont {Klavins}, \citenamefont {Cooley}, \citenamefont
  {Yang},\ and\ \citenamefont {Curro}}]{ShirerPNAS2012}%
  \BibitemOpen
  \bibfield  {author} {\bibinfo {author} {\bibfnamefont {Kent~R.}\ \bibnamefont
  {Shirer}}, \bibinfo {author} {\bibfnamefont {Abigail~C.}\ \bibnamefont
  {Shockley}}, \bibinfo {author} {\bibfnamefont {Adam~P.}\ \bibnamefont
  {Dioguardi}}, \bibinfo {author} {\bibfnamefont {John}\ \bibnamefont
  {Crocker}}, \bibinfo {author} {\bibfnamefont {Ching~H.}\ \bibnamefont {Lin}},
  \bibinfo {author} {\bibfnamefont {Nicholas}\ \bibnamefont {apRoberts
  Warren}}, \bibinfo {author} {\bibfnamefont {David~M.}\ \bibnamefont
  {Nisson}}, \bibinfo {author} {\bibfnamefont {Peter}\ \bibnamefont {Klavins}},
  \bibinfo {author} {\bibfnamefont {Jason~C.}\ \bibnamefont {Cooley}}, \bibinfo
  {author} {\bibfnamefont {Yi-feng}\ \bibnamefont {Yang}}, \ and\ \bibinfo
  {author} {\bibfnamefont {Nicholas~J.}\ \bibnamefont {Curro}},\ }\bibfield
  {title} {\enquote {\bibinfo {title} {Long range order and two-fluid behavior
  in heavy electron materials},}\ }\href {\doibase 10.1073/pnas.1209609109}
  {\bibfield  {journal} {\bibinfo  {journal} {Proc. Natl. Acad. Sci.}\ }\textbf
  {\bibinfo {volume} {109}},\ \bibinfo {pages} {E3067--E3073} (\bibinfo {year}
  {2012})}\BibitemShut {NoStop}%
\bibitem [{\citenamefont {Curro}\ \emph {et~al.}(2001)\citenamefont {Curro},
  \citenamefont {Simovic}, \citenamefont {Hammel}, \citenamefont {Pagliuso},
  \citenamefont {Sarrao}, \citenamefont {Thompson},\ and\ \citenamefont
  {Martins}}]{Curro2001}%
  \BibitemOpen
  \bibfield  {author} {\bibinfo {author} {\bibfnamefont {NJ}~\bibnamefont
  {Curro}}, \bibinfo {author} {\bibfnamefont {B}~\bibnamefont {Simovic}},
  \bibinfo {author} {\bibfnamefont {PC}~\bibnamefont {Hammel}}, \bibinfo
  {author} {\bibfnamefont {PG}~\bibnamefont {Pagliuso}}, \bibinfo {author}
  {\bibfnamefont {JL}~\bibnamefont {Sarrao}}, \bibinfo {author} {\bibfnamefont
  {JD}~\bibnamefont {Thompson}}, \ and\ \bibinfo {author} {\bibfnamefont
  {GB}~\bibnamefont {Martins}},\ }\bibfield  {title} {\enquote {\bibinfo
  {title} {Anomalous {NMR} magnetic shifts in {CeCoIn$_5$}},}\ }\href {\doibase
  10.1103/PhysRevB.64.180514} {\bibfield  {journal} {\bibinfo  {journal} {Phys.
  Rev. B}\ }\textbf {\bibinfo {volume} {64}},\ \bibinfo {pages} {180514(R)}
  (\bibinfo {year} {2001})}\BibitemShut {NoStop}%
\bibitem [{\citenamefont {Yang}\ \emph {et~al.}(2009)\citenamefont {Yang},
  \citenamefont {Urbano}, \citenamefont {Curro},\ and\ \citenamefont
  {Pines}}]{Yang2009a}%
  \BibitemOpen
  \bibfield  {author} {\bibinfo {author} {\bibfnamefont {Y.-F.}\ \bibnamefont
  {Yang}}, \bibinfo {author} {\bibfnamefont {Ricardo}\ \bibnamefont {Urbano}},
  \bibinfo {author} {\bibfnamefont {Nicholas~J.}\ \bibnamefont {Curro}}, \ and\
  \bibinfo {author} {\bibfnamefont {David}\ \bibnamefont {Pines}},\ }\bibfield
  {title} {\enquote {\bibinfo {title} {Magnetic excitations in {K}ondo liquid:
  Superconductivity and hidden magnetic quantum critical fluctuations},}\
  }\href {\doibase 10.1103/PhysRevLett.103.197004} {\bibfield  {journal}
  {\bibinfo  {journal} {Phys. Rev. Lett.}\ }\textbf {\bibinfo {volume} {103}},\
  \bibinfo {pages} {197004} (\bibinfo {year} {2009})}\BibitemShut {NoStop}%
\bibitem [{\citenamefont {Slichter}(1992)}]{CPSbook}%
  \BibitemOpen
  \bibfield  {author} {\bibinfo {author} {\bibfnamefont {C.~P.}\ \bibnamefont
  {Slichter}},\ }\href@noop {} {\emph {\bibinfo {title} {Principles of Nuclear
  Magnetic Resonance}}},\ \bibinfo {edition} {3rd}\ ed.\ (\bibinfo  {publisher}
  {Springer-Verlag},\ \bibinfo {year} {1992})\BibitemShut {NoStop}%
\bibitem [{\citenamefont {Shockley}\ \emph {et~al.}(2012)\citenamefont
  {Shockley}, \citenamefont {Dioguardi}, \citenamefont {apRoberts Warren},
  \citenamefont {Klavins}, \citenamefont {Capan}, \citenamefont {Fisk},\ and\
  \citenamefont {Curro}}]{Shockley2012}%
  \BibitemOpen
  \bibfield  {author} {\bibinfo {author} {\bibfnamefont {A.~C.}\ \bibnamefont
  {Shockley}}, \bibinfo {author} {\bibfnamefont {A.~P.}\ \bibnamefont
  {Dioguardi}}, \bibinfo {author} {\bibfnamefont {N.}~\bibnamefont {apRoberts
  Warren}}, \bibinfo {author} {\bibfnamefont {P.}~\bibnamefont {Klavins}},
  \bibinfo {author} {\bibfnamefont {C.}~\bibnamefont {Capan}}, \bibinfo
  {author} {\bibfnamefont {Z.}~\bibnamefont {Fisk}}, \ and\ \bibinfo {author}
  {\bibfnamefont {N.~J.}\ \bibnamefont {Curro}},\ }\bibfield  {title} {\enquote
  {\bibinfo {title} {Investigating the structure of {Ce$_{1-x}$La$_x$CoIn$_5$}
  using {NQR}},}\ }\href {\doibase 10.1007/s10948-012-1639-5} {\bibfield
  {journal} {\bibinfo  {journal} {J. Supercond. Novel Magn.}\ }\textbf
  {\bibinfo {volume} {25}},\ \bibinfo {pages} {2141 -- 2144} (\bibinfo {year}
  {2012})}\BibitemShut {NoStop}%
\bibitem [{\citenamefont {Curro}\ \emph {et~al.}(2004)\citenamefont {Curro},
  \citenamefont {Young}, \citenamefont {Schmalian},\ and\ \citenamefont
  {Pines}}]{Curro2004}%
  \BibitemOpen
  \bibfield  {author} {\bibinfo {author} {\bibfnamefont {NJ}~\bibnamefont
  {Curro}}, \bibinfo {author} {\bibfnamefont {BL}~\bibnamefont {Young}},
  \bibinfo {author} {\bibfnamefont {J}~\bibnamefont {Schmalian}}, \ and\
  \bibinfo {author} {\bibfnamefont {D}~\bibnamefont {Pines}},\ }\bibfield
  {title} {\enquote {\bibinfo {title} {Scaling in the emergent behavior of
  heavy-electron materials},}\ }\href {\doibase 10.1103/PhysRevB.70.235117}
  {\bibfield  {journal} {\bibinfo  {journal} {Phys. Rev. B}\ }\textbf {\bibinfo
  {volume} {70}},\ \bibinfo {pages} {235117} (\bibinfo {year}
  {2004})}\BibitemShut {NoStop}%
\bibitem [{\citenamefont {Jiang}\ \emph {et~al.}(2014)\citenamefont {Jiang},
  \citenamefont {Curro},\ and\ \citenamefont {Scalettar}}]{jiang14}%
  \BibitemOpen
  \bibfield  {author} {\bibinfo {author} {\bibfnamefont {M.}~\bibnamefont
  {Jiang}}, \bibinfo {author} {\bibfnamefont {N.~J.}\ \bibnamefont {Curro}}, \
  and\ \bibinfo {author} {\bibfnamefont {R.~T.}\ \bibnamefont {Scalettar}},\
  }\bibfield  {title} {\enquote {\bibinfo {title} {Universal {K}night shift
  anomaly in the periodic {A}nderson model},}\ }\href {\doibase
  10.1103/PhysRevB.90.241109} {\bibfield  {journal} {\bibinfo  {journal} {Phys.
  Rev. B}\ }\textbf {\bibinfo {volume} {90}},\ \bibinfo {pages} {241109}
  (\bibinfo {year} {2014})}\BibitemShut {NoStop}%
\bibitem [{\citenamefont {Lin}\ \emph {et~al.}(2015)\citenamefont {Lin},
  \citenamefont {Shirer}, \citenamefont {Crocker}, \citenamefont {Dioguardi},
  \citenamefont {Lawson}, \citenamefont {Bush}, \citenamefont {Klavins},\ and\
  \citenamefont {Curro}}]{Lin2015}%
  \BibitemOpen
  \bibfield  {author} {\bibinfo {author} {\bibfnamefont {C.~H.}\ \bibnamefont
  {Lin}}, \bibinfo {author} {\bibfnamefont {K.~R.}\ \bibnamefont {Shirer}},
  \bibinfo {author} {\bibfnamefont {J.}~\bibnamefont {Crocker}}, \bibinfo
  {author} {\bibfnamefont {A.~P.}\ \bibnamefont {Dioguardi}}, \bibinfo {author}
  {\bibfnamefont {M.~M.}\ \bibnamefont {Lawson}}, \bibinfo {author}
  {\bibfnamefont {B.~T.}\ \bibnamefont {Bush}}, \bibinfo {author}
  {\bibfnamefont {P.}~\bibnamefont {Klavins}}, \ and\ \bibinfo {author}
  {\bibfnamefont {N.~J.}\ \bibnamefont {Curro}},\ }\bibfield  {title} {\enquote
  {\bibinfo {title} {Evolution of hyperfine parameters across a quantum
  critical point in {${\mathrm{CeRhIn}}_{5}$}},}\ }\href {\doibase
  10.1103/PhysRevB.92.155147} {\bibfield  {journal} {\bibinfo  {journal} {Phys.
  Rev. B}\ }\textbf {\bibinfo {volume} {92}},\ \bibinfo {pages} {155147}
  (\bibinfo {year} {2015})}\BibitemShut {NoStop}%
\bibitem [{\citenamefont {Park}\ \emph {et~al.}(2006)\citenamefont {Park},
  \citenamefont {Ronning}, \citenamefont {Yuan}, \citenamefont {Salamon},
  \citenamefont {Movshovich}, \citenamefont {Sarrao},\ and\ \citenamefont
  {Thompson}}]{tuson}%
  \BibitemOpen
  \bibfield  {author} {\bibinfo {author} {\bibfnamefont {T.}~\bibnamefont
  {Park}}, \bibinfo {author} {\bibfnamefont {F.}~\bibnamefont {Ronning}},
  \bibinfo {author} {\bibfnamefont {H.~Q.}\ \bibnamefont {Yuan}}, \bibinfo
  {author} {\bibfnamefont {M.~B.}\ \bibnamefont {Salamon}}, \bibinfo {author}
  {\bibfnamefont {R.}~\bibnamefont {Movshovich}}, \bibinfo {author}
  {\bibfnamefont {J.~L.}\ \bibnamefont {Sarrao}}, \ and\ \bibinfo {author}
  {\bibfnamefont {J.~D.}\ \bibnamefont {Thompson}},\ }\bibfield  {title}
  {\enquote {\bibinfo {title} {Hidden magnetism and quantum criticality in the
  heavy fermion superconductor {CeRhIn$_5$}},}\ }\href {\doibase
  10.1038/nature04571} {\bibfield  {journal} {\bibinfo  {journal} {Nature}\
  }\textbf {\bibinfo {volume} {440}},\ \bibinfo {pages} {65 -- 68} (\bibinfo
  {year} {2006})}\BibitemShut {NoStop}%
\bibitem [{\citenamefont {Yang}\ and\ \citenamefont
  {Pines}(2008)}]{YangDavidPRL}%
  \BibitemOpen
  \bibfield  {author} {\bibinfo {author} {\bibfnamefont {Y.-F.}\ \bibnamefont
  {Yang}}\ and\ \bibinfo {author} {\bibfnamefont {D.}~\bibnamefont {Pines}},\
  }\bibfield  {title} {\enquote {\bibinfo {title} {Universal behavior in
  heavy-electron materials},}\ }\href {\doibase 10.1103/PhysRevLett.100.096404}
  {\bibfield  {journal} {\bibinfo  {journal} {Phys. Rev. Lett.}\ }\textbf
  {\bibinfo {volume} {100}},\ \bibinfo {eid} {096404} (\bibinfo {year}
  {2008})}\BibitemShut {NoStop}%
\bibitem [{\citenamefont {Bauer}\ \emph {et~al.}(2011)\citenamefont {Bauer},
  \citenamefont {Yang}, \citenamefont {Capan}, \citenamefont {Urbano},
  \citenamefont {Miclea}, \citenamefont {Sakai}, \citenamefont {Ronning},
  \citenamefont {Graf}, \citenamefont {Balatsky}, \citenamefont {Movshovich},
  \citenamefont {Bianchi}, \citenamefont {Reyes}, \citenamefont {Kuhns},
  \citenamefont {Thompson},\ and\ \citenamefont {Fisk}}]{Bauer2011}%
  \BibitemOpen
  \bibfield  {author} {\bibinfo {author} {\bibfnamefont {E.~D.}\ \bibnamefont
  {Bauer}}, \bibinfo {author} {\bibfnamefont {Yi-feng}\ \bibnamefont {Yang}},
  \bibinfo {author} {\bibfnamefont {C.}~\bibnamefont {Capan}}, \bibinfo
  {author} {\bibfnamefont {R.~R.}\ \bibnamefont {Urbano}}, \bibinfo {author}
  {\bibfnamefont {C.~F.}\ \bibnamefont {Miclea}}, \bibinfo {author}
  {\bibfnamefont {H.}~\bibnamefont {Sakai}}, \bibinfo {author} {\bibfnamefont
  {F.}~\bibnamefont {Ronning}}, \bibinfo {author} {\bibfnamefont {M.~J.}\
  \bibnamefont {Graf}}, \bibinfo {author} {\bibfnamefont {A.~V.}\ \bibnamefont
  {Balatsky}}, \bibinfo {author} {\bibfnamefont {R.}~\bibnamefont
  {Movshovich}}, \bibinfo {author} {\bibfnamefont {A.~D.}\ \bibnamefont
  {Bianchi}}, \bibinfo {author} {\bibfnamefont {A.~P.}\ \bibnamefont {Reyes}},
  \bibinfo {author} {\bibfnamefont {P.~L.}\ \bibnamefont {Kuhns}}, \bibinfo
  {author} {\bibfnamefont {J.~D.}\ \bibnamefont {Thompson}}, \ and\ \bibinfo
  {author} {\bibfnamefont {Z.}~\bibnamefont {Fisk}},\ }\bibfield  {title}
  {\enquote {\bibinfo {title} {Electronic inhomogeneity in a {K}ondo
  lattice},}\ }\href {\doibase 10.1073/pnas.1103965108} {\bibfield  {journal}
  {\bibinfo  {journal} {Proc. Natl. Acad. Sci.}\ }\textbf {\bibinfo {volume}
  {108}},\ \bibinfo {pages} {6857--6861} (\bibinfo {year} {2011})}\BibitemShut
  {NoStop}%
\bibitem [{\citenamefont {Seo}\ \emph {et~al.}(2014)\citenamefont {Seo},
  \citenamefont {Lu}, \citenamefont {Zhu}, \citenamefont {Urbano},
  \citenamefont {Curro}, \citenamefont {Bauer}, \citenamefont {Sidorov},
  \citenamefont {Pham}, \citenamefont {Park}, \citenamefont {Fisk},\ and\
  \citenamefont {Thompson}}]{ParkDropletsNature2013}%
  \BibitemOpen
  \bibfield  {author} {\bibinfo {author} {\bibfnamefont {S.}~\bibnamefont
  {Seo}}, \bibinfo {author} {\bibfnamefont {Xin}\ \bibnamefont {Lu}}, \bibinfo
  {author} {\bibfnamefont {J-X.}\ \bibnamefont {Zhu}}, \bibinfo {author}
  {\bibfnamefont {R.~R.}\ \bibnamefont {Urbano}}, \bibinfo {author}
  {\bibfnamefont {N.}~\bibnamefont {Curro}}, \bibinfo {author} {\bibfnamefont
  {E.~D.}\ \bibnamefont {Bauer}}, \bibinfo {author} {\bibfnamefont {V.~A.}\
  \bibnamefont {Sidorov}}, \bibinfo {author} {\bibfnamefont {L.~D.}\
  \bibnamefont {Pham}}, \bibinfo {author} {\bibfnamefont {Tuson}\ \bibnamefont
  {Park}}, \bibinfo {author} {\bibfnamefont {Z.}~\bibnamefont {Fisk}}, \ and\
  \bibinfo {author} {\bibfnamefont {J.~D.}\ \bibnamefont {Thompson}},\
  }\bibfield  {title} {\enquote {\bibinfo {title} {Disorder in quantum critical
  superconductors},}\ }\href {\doibase 10.1038/nphys2820} {\bibfield  {journal}
  {\bibinfo  {journal} {Nat. Phys.}\ }\textbf {\bibinfo {volume} {10}},\
  \bibinfo {pages} {120--125} (\bibinfo {year} {2014})}\BibitemShut {NoStop}%
\bibitem [{\citenamefont {S\o{}rensen}\ and\ \citenamefont
  {Affleck}(1996)}]{AffleckKondoScreening}%
  \BibitemOpen
  \bibfield  {author} {\bibinfo {author} {\bibfnamefont {Erik~S.}\ \bibnamefont
  {S\o{}rensen}}\ and\ \bibinfo {author} {\bibfnamefont {Ian}\ \bibnamefont
  {Affleck}},\ }\bibfield  {title} {\enquote {\bibinfo {title} {Scaling theory
  of the {K}ondo screening cloud},}\ }\href {\doibase 10.1103/PhysRevB.53.9153}
  {\bibfield  {journal} {\bibinfo  {journal} {Phys. Rev. B}\ }\textbf {\bibinfo
  {volume} {53}},\ \bibinfo {pages} {9153--9167} (\bibinfo {year}
  {1996})}\BibitemShut {NoStop}%
\bibitem [{\citenamefont {Mila}\ and\ \citenamefont {Rice}(1989)}]{mila89}%
  \BibitemOpen
  \bibfield  {author} {\bibinfo {author} {\bibfnamefont {F.}~\bibnamefont
  {Mila}}\ and\ \bibinfo {author} {\bibfnamefont {T.M}\ \bibnamefont {Rice}},\
  }\bibfield  {title} {\enquote {\bibinfo {title} {Analysis of magnetic
  resonance experiments in {YBa$_2$Cu$_3$O$_7$}},}\ }\href {\doibase
  10.1016/0921-4534(89)90286-4} {\bibfield  {journal} {\bibinfo  {journal}
  {Physica}\ }\textbf {\bibinfo {volume} {157C}},\ \bibinfo {pages} {561}
  (\bibinfo {year} {1989})}\BibitemShut {NoStop}%
\bibitem [{\citenamefont {Abragam}(1961)}]{abragambook}%
  \BibitemOpen
  \bibfield  {author} {\bibinfo {author} {\bibfnamefont {A.}~\bibnamefont
  {Abragam}},\ }\href@noop {} {\emph {\bibinfo {title} {The Principles of
  Nuclear Magnetism}}}\ (\bibinfo  {publisher} {Oxford University Press,
  Oxford},\ \bibinfo {year} {1961})\BibitemShut {NoStop}%
\bibitem [{\citenamefont {Benali}\ \emph {et~al.}(2016)\citenamefont {Benali},
  \citenamefont {Bai}, \citenamefont {Curro},\ and\ \citenamefont
  {Scalettar}}]{BenAliPAMdopants2016}%
  \BibitemOpen
  \bibfield  {author} {\bibinfo {author} {\bibfnamefont {A.}~\bibnamefont
  {Benali}}, \bibinfo {author} {\bibfnamefont {Z.~J.}\ \bibnamefont {Bai}},
  \bibinfo {author} {\bibfnamefont {N.~J.}\ \bibnamefont {Curro}}, \ and\
  \bibinfo {author} {\bibfnamefont {R.~T.}\ \bibnamefont {Scalettar}},\
  }\bibfield  {title} {\enquote {\bibinfo {title} {Impurity-induced
  antiferromagnetic domains in the periodic anderson model},}\ }\href {\doibase
  10.1103/PhysRevB.94.085132} {\bibfield  {journal} {\bibinfo  {journal} {Phys.
  Rev. B}\ }\textbf {\bibinfo {volume} {94}},\ \bibinfo {pages} {085132}
  (\bibinfo {year} {2016})}\BibitemShut {NoStop}%
\bibitem [{\citenamefont {Kohori}\ \emph {et~al.}(2001)\citenamefont {Kohori},
  \citenamefont {Yamato}, \citenamefont {Iwamoto}, \citenamefont {Kohara},
  \citenamefont {Bauer}, \citenamefont {Maple},\ and\ \citenamefont
  {Sarrao}}]{kohori115shift}%
  \BibitemOpen
  \bibfield  {author} {\bibinfo {author} {\bibfnamefont {Y.}~\bibnamefont
  {Kohori}}, \bibinfo {author} {\bibfnamefont {Y.}~\bibnamefont {Yamato}},
  \bibinfo {author} {\bibfnamefont {Y.}~\bibnamefont {Iwamoto}}, \bibinfo
  {author} {\bibfnamefont {T.}~\bibnamefont {Kohara}}, \bibinfo {author}
  {\bibfnamefont {E.~D.}\ \bibnamefont {Bauer}}, \bibinfo {author}
  {\bibfnamefont {M.~B.}\ \bibnamefont {Maple}}, \ and\ \bibinfo {author}
  {\bibfnamefont {J.~L.}\ \bibnamefont {Sarrao}},\ }\bibfield  {title}
  {\enquote {\bibinfo {title} {{NMR} and {NQR} studies of the heavy fermion
  superconductors {CeTIn$_5 $ (T=Co and Ir)}},}\ }\href {\doibase
  10.1103/PhysRevB.64.134526} {\bibfield  {journal} {\bibinfo  {journal} {Phys.
  Rev. B}\ }\textbf {\bibinfo {volume} {64}},\ \bibinfo {pages} {134526}
  (\bibinfo {year} {2001})}\BibitemShut {NoStop}%
\bibitem [{\citenamefont {Urbano}\ \emph {et~al.}(2007)\citenamefont {Urbano},
  \citenamefont {Young}, \citenamefont {Curro}, \citenamefont {Thompson},
  \citenamefont {Pham},\ and\ \citenamefont {Fisk}}]{Urbano2007}%
  \BibitemOpen
  \bibfield  {author} {\bibinfo {author} {\bibfnamefont {R.~R.}\ \bibnamefont
  {Urbano}}, \bibinfo {author} {\bibfnamefont {B.-L.}\ \bibnamefont {Young}},
  \bibinfo {author} {\bibfnamefont {N.~J.}\ \bibnamefont {Curro}}, \bibinfo
  {author} {\bibfnamefont {J.~D.}\ \bibnamefont {Thompson}}, \bibinfo {author}
  {\bibfnamefont {L.~D.}\ \bibnamefont {Pham}}, \ and\ \bibinfo {author}
  {\bibfnamefont {Z.}~\bibnamefont {Fisk}},\ }\bibfield  {title} {\enquote
  {\bibinfo {title} {Interacting antiferromagnetic droplets in quantum critical
  {CeCoIn$_5$}},}\ }\href {\doibase 10.1103/PhysRevLett.99.146402} {\bibfield
  {journal} {\bibinfo  {journal} {Phys. Rev. Lett.}\ }\textbf {\bibinfo
  {volume} {99}},\ \bibinfo {pages} {146402} (\bibinfo {year}
  {2007})}\BibitemShut {NoStop}%
\bibitem [{\citenamefont {Si}\ \emph {et~al.}(2001)\citenamefont {Si},
  \citenamefont {Rabello}, \citenamefont {Ingersent},\ and\ \citenamefont
  {Smith}}]{SiLocalQCP}%
  \BibitemOpen
  \bibfield  {author} {\bibinfo {author} {\bibfnamefont {Q.~M.}\ \bibnamefont
  {Si}}, \bibinfo {author} {\bibfnamefont {S.}~\bibnamefont {Rabello}},
  \bibinfo {author} {\bibfnamefont {K.}~\bibnamefont {Ingersent}}, \ and\
  \bibinfo {author} {\bibfnamefont {J.~L.}\ \bibnamefont {Smith}},\ }\bibfield
  {title} {\enquote {\bibinfo {title} {Locally critical quantum phase
  transitions in strongly correlated metals},}\ }\href {\doibase
  10.1038/35101507} {\bibfield  {journal} {\bibinfo  {journal} {Nature}\
  }\textbf {\bibinfo {volume} {413}},\ \bibinfo {pages} {804 -- 808} (\bibinfo
  {year} {2001})}\BibitemShut {NoStop}%
\bibitem [{\citenamefont {Schroder}\ \emph {et~al.}(2000)\citenamefont
  {Schroder}, \citenamefont {Aeppli}, \citenamefont {Coldea}, \citenamefont
  {Adams}, \citenamefont {Stockert}, \citenamefont {Lohneysen}, \citenamefont
  {Bucher}, \citenamefont {Ramazashvili},\ and\ \citenamefont
  {Coleman}}]{Schroder2000}%
  \BibitemOpen
  \bibfield  {author} {\bibinfo {author} {\bibfnamefont {A.}~\bibnamefont
  {Schroder}}, \bibinfo {author} {\bibfnamefont {G.}~\bibnamefont {Aeppli}},
  \bibinfo {author} {\bibfnamefont {R.}~\bibnamefont {Coldea}}, \bibinfo
  {author} {\bibfnamefont {M.}~\bibnamefont {Adams}}, \bibinfo {author}
  {\bibfnamefont {O.}~\bibnamefont {Stockert}}, \bibinfo {author}
  {\bibfnamefont {H.v.}\ \bibnamefont {Lohneysen}}, \bibinfo {author}
  {\bibfnamefont {E.}~\bibnamefont {Bucher}}, \bibinfo {author} {\bibfnamefont
  {R.}~\bibnamefont {Ramazashvili}}, \ and\ \bibinfo {author} {\bibfnamefont
  {P.}~\bibnamefont {Coleman}},\ }\bibfield  {title} {\enquote {\bibinfo
  {title} {Onset of antiferromagnetism in heavy-fermion metals},}\ }\href
  {http://dx.doi.org/10.1038/35030039} {\bibfield  {journal} {\bibinfo
  {journal} {Nature}\ }\textbf {\bibinfo {volume} {407}},\ \bibinfo {pages}
  {351--355} (\bibinfo {year} {2000})}\BibitemShut {NoStop}%
\bibitem [{\citenamefont {ying Wei}\ and\ \citenamefont {feng
  Yang}(2017)}]{Wei2017}%
  \BibitemOpen
  \bibfield  {author} {\bibinfo {author} {\bibfnamefont {Lan}\ \bibnamefont
  {ying Wei}}\ and\ \bibinfo {author} {\bibfnamefont {Yi}~\bibnamefont {feng
  Yang}},\ }\bibfield  {title} {\enquote {\bibinfo {title} {Doping-induced
  perturbation and percolation in the two-dimensional anderson lattice},}\
  }\href {\doibase 10.1038/srep46089} {\bibfield  {journal} {\bibinfo
  {journal} {Scientific Reports}\ }\textbf {\bibinfo {volume} {7}},\ \bibinfo
  {pages} {46089} (\bibinfo {year} {2017})}\BibitemShut {NoStop}%
\end{thebibliography}%

\end{document}